\documentclass[preprint, 10pt, authoryear]{elsarticle}
\usepackage{amsmath,amssymb,amsthm,amsfonts}
\usepackage{mathabx}
\usepackage{mathrsfs}
\usepackage{bm}
\usepackage{graphicx}
\usepackage{float}
\usepackage{subfig}
\usepackage{color,xcolor}
\usepackage{tabularx}
\usepackage{epsfig,psfrag}
\usepackage{tikz}
\usepackage{pgflibraryshapes}
\usepackage{graphicx,psfrag}
\setcitestyle{authoryear,round}
\usepackage{hyperref}

\usepackage{changes}
\makeatletter
\@namedef{Changes@AuthorColor}{blue}
\colorlet{Changes@Color}{blue}
\makeatother

\makeatletter
\hypersetup{colorlinks=true}
\AtBeginDocument{\@ifpackageloaded{hyperref}
	{\def\@linkcolor{magenta}
		\def\@anchorcolor{black}
		\def\@citecolor{teal}
		\def\@filecolor{cyan}
		\def\@urlcolor{magenta}
		\def\@menucolor{red}
		\def\@pagecolor{cyan}
		\begingroup
		\@makeother\`%
		\@makeother\=%
		\edef\x{%
			\edef\noexpand\x{%
				\endgroup
				\noexpand\toks@{%
					\catcode 96=\noexpand\the\catcode`\noexpand\`\relax
					\catcode 61=\noexpand\the\catcode`\noexpand\=\relax
				}%
			}%
			\noexpand\x
		}%
		\x
		\@makeother\`
		\@makeother\=
	}{}}
\makeatother

\makeatletter \oddsidemargin 0.0cm \evensidemargin 0.0cm
\newcommand{\be}{\begin{equation}}
\newcommand{\en}{\end{equation}}
\newcommand{\la}{\label}
\newcommand{\ep}{{\varepsilon}}

\def\rr#1{(\ref{#1})}
\def\bm#1{\mbox{\boldmath{$#1$}}}

\newcommand{\s}[1]{{\Large\textsf{\textbf{#1}}}}

\numberwithin{equation}{section}
\theoremstyle{plain}

\newtheorem{theorem*}{Theorem}

\theoremstyle{definition}

\journal{Journal of the Mechanics and Physics of Solids}

\begin{document}

\begin{frontmatter}

\title{\s{Analysis of axisymmetric necking of a circular dielectric membrane based on a one-dimensional model}}

\author[mymainaddress]{Xiang Yu\corref{mycorrespondingauthor}}
\cortext[mycorrespondingauthor]{Corresponding author}
\ead{yuxiang@dgut.edu.cn}
\author[mysecondaryaddress]{Yibin Fu}

\address[mymainaddress]{Department of Mathematics, School of Computer Science and Technology, Dongguan University of Technology, Dongguan, 523808, China}

\address[mysecondaryaddress]{School of Computer Science and Mathematics, Keele University, Staffs ST5 5BG, UK}

\begin{abstract}

To facilitate the understanding of the mechanisms underlying the electric breakdown of dielectric elastomers, we derive a one-dimensional (1d) model for axisymmetric necking in a dielectric membrane subjected to equibiaxial stretching and an electric field, starting from the three-dimensional (3d) nonlinear electroelasticity theory.  Our reduction is built on the variational asymptotic method, so that the resulting 1d model is asymptotically self-consistent. The 1d model offers an easier and more efficient way to analyze axisymmetric necking in a dielectric membrane in the linear, weakly nonlinear and fully nonlinear regimes. It delivers results identical to the 3d theory in the linear and weakly nonlinear regimes, and near-identical results in the fully nonlinear regime due to its asymptotic self-consistency. We demonstrate the straightforward implementation of this 1d model by solving it using the Rayleigh--Ritz method and validate it by comparison with finite-element simulations. The 1d model enables a precise calculation of the minimum thickness that a dielectric membrane can reach when
necking instability occurs and quantitative assessment of the effects of imperfections so that the integrity of a dielectric elastomer actuator can be evaluated with respect to electric breakdown. The developed methodology is not problem-specific and can also be applied to analyze similar phenomena in other soft materials subjected to any fields (e.g., the axisymmetric necking of stretched plastic membranes).
\end{abstract}

\begin{keyword}
Nonlinear electroleasticity\sep Dielectric membranes\sep Stability and bifurcation \sep Localization \sep Variational asymptotic method
\end{keyword}
\end{frontmatter}

\section{Introduction}

Dielectric elastomers are a newly emerging class of electroactive materials that is capable of large deformations under the actuation of an applied voltage. They have the advantages of fast response, high energy density, light weight, good flexibility, etc., and therefore have promising applications in areas like soft robotics \citep{pelrine1998electrostriction,pelrine2000high,li2021self,yarali2022magneto}, stretchable electronics \citep{rogers2010materials,shi2018dielectric,shi2022processable}, energy harvesting \citep{kornbluh2012dielectric,yang2017avoiding} and optics \citep{carpi2011bioinspired}. However, the large deformation process of dielectric elastomers is often accompanied by various electromechanical instabilities.
The two predominant types of instabilities observed in dielectric elastomer devices are the pull-in  and wrinkling instabilities.
The pull-in instability refers to a sudden thinning deformation that often preludes electric breakdown and signifies the failure of devices \citep{stark1955electric,blok1969dielectric,plante2006large,zhao2007method,zhao2009electromechanical,zhang2011mechanical,huang2012electromechanical,
de2013electromechanical,de2013inhomogeneous,de2014failure,zurlo2017catastrophic,yang2022inhomogeneous}, whereas the wrinkling instability involves relieving in-plane compression through out-of-plane deformations \citep{bertoldi2011instabilities,rudykh2011stability,dorfmann2014instabilities,gei2014role,yang2017revisiting,su2018wrinkles,dorfmann2019instabilities,greaney2019out,broderick2020stability,su2020voltage,su2020effect,xia2021instability,bahreman2022structural,khurana2022electromechanical}. These two types of electromechanical instabilities have received extensive attention over the past two decades and significant progress has been made. The reader is referred to \cite{dorfmann2017nonlinear} and \cite{lu2020mechanics} for a comprehensive review of the relevant literature. The Treloar-Kearsley instability \citep{kearsley1986asymmetric}, whereby unequal stretches bifurcate from an equibiaixal stretching state, may also occur, at least theoretically, and its use to maximize in-plane deformations has been explored by \cite{yang2017avoiding}.

The rapid thinning and subsequent breakdown of dielectric films  was first studied by \cite{stark1955electric}, where the authors proposed that the rapid thinning is due to the electric field reaching a maximum value and explored the connection between this critical value and Young's modulus using a simple linear theory. The critical electric field predicted by the Stark--Garton model was found in excellent numerical agreement with experiment data for polyisobutylene. On the other hand, the model overestimated the electric breakdown fields for polystyrene and polymethylmethacrylate \citep{zhou2009electrical}. One possible reason for this inconsistency,  as noted by \cite{blok1969dielectric}, is that the Stark--Garton model assumed that the dielectric deforms homogeneously under the electric field.
In practice, deforming the entire area of a dielectric requires immense stress and is thus difficult to achieve.
\cite{blok1969dielectric} suggested that  a localized thinning, or necking stated differently, would occur at a critical electric field, as a result of the intrinsic inhomogeneity of the electric field and dielectric material. The thin region experiences a higher electric field than the thick region under the same voltage, resulting in a much lower  electric breakdown strength than that predicted by the Stark--Garton model. The necking phenomenon of dielectric films was later observed experimentally by  \cite{plante2006large} and simulated numerically by \cite{zhou2008propagation}. Inspired by the work of \cite{blok1969dielectric},  Zurlo and coworkers \citep{de2013electromechanical,de2013inhomogeneous,de2014failure,zurlo2017catastrophic,greaney2019out} analyzed localized necking in a dielectric membrane
with the aid of an approximate model based on a non-asymptotic expansion of the displacement field. This approximate model was further assessed in \cite{fu2018reduced}. For the case of simple tension in a state of plane strain, \cite{fu2018localized} investigated localized necking in a dielectric membrane using an analogy with the inflation problem for a rubber tube \citep{fu2008post}. It was shown that localized necking would initiate when the Hessian determinant of the free energy function became zero, and would quickly evolve into a \lq\lq two-phase'' deformation observed by \cite{plante2006large} and analyzed by \cite{zhao2007electromechanical} and \cite{huang2012electromechanical}.

The current study on {\it axisymmetric} necking may be viewed as a sequel to two  previous studies: \cite{wang2022axisymmetric} and \cite{fu2023axisymmetric} that are concerned with the purely mechanical and the coupled electroelastic cases, respectively. In the study by \cite{wang2022axisymmetric}, the bifurcation condition for axisymmetric necking was derived and was shown to not correspond to the vanishing of the Hessian determinant of the free energy function. A weakly nonlinear analysis was also conducted to derive an amplitude equation in the form of a fourth-order nonlinear ordinary differential equation with variable coefficients. Without being able to solve this amplitude equation analytically, the authors resorted to Abaqus simulations in order to demonstrate the initiation of a localized necking solution and its further evolution. In the subsequent study by \cite{fu2023axisymmetric}, the above amplitude equation was solved by the finite difference method and extensions were then made to the electroelastic case. However, although Abaqus simulations can be used to investigate the fully nonlinear regime, their implementations can be cumbersome, especially for the coupled electroelastic case. This provides the main motivation for the current study, which can also be viewed as an extension of the study by \cite{audoly2016analysis}, in terms of methodology, from the prismatic case (where necking occurs in the axial direction) to the membrane case (where necking occurs in the radial direction). The same methodology has been applied to derive approximate models with remarkable accuracy for a variety of other localization problems, including  elastocapillary necking \citep{lestringant2020one}, bulging in rubber tubes \citep{lestringant2018diffuse,yu2023one}, and morphoelastic rods and shells \citep{moulton2020morphoelastic,kaczmarski2022active,yu2024asymptotically}. The methodology was further abstracted by \cite{lestringant2020asymptotically} into a systematic and non-problem-specific algorithm so that a reduced model to any asymptotic order for any given problem can be derived following the same generic procedure. We note that  this methodology shares the same philosophy as the variational asymptotic method pioneered  earlier  by \cite{berdichevskii1979variational}, see also \cite{berdichevsky2009variational}, although the variational asymptotic method was originally developed for small-strain problems. It has been applied to derive reduced theories for beams  \citep{yu-hodges-2004}, plates  \citep{hodges-atilgan-danielson-1993} and shells  \citep{libai-simmonds-1983, le1999},  and to derive homogenised material properties of periodic structures  \citep{yu-tang-2007}. Thus, for easy reference, we shall from now on adopt the term ``variational asymptotic method".

In this work, we derive a 1d model for axisymmetric necking in a dielectric membrane under equibiaxial stretching using the variational asymptotic method. Although a translational invariance is missing in the direction of localization, it will be shown that a 1d reduced model can still be derived from  the 3d nonlinear electroelastic model, with the associated energy functional
\begin{align}\label{eq:1d-introd}
\begin{split}
\mathcal{E}_{\text{1d}}[\mu]&=\int_0^A \Big(G(\rho,\mu) +\frac{H^2}{24}\frac{G_1(\rho,\mu)}{\rho}\lambda'(R)^2\Big)R\,dR-P A^2\mu(A),
\end{split}
\end{align}
where $A$ is the undeformed radius and thickness of a circular membrane, $H$ is its undeformed thickness, $R$ is the radial coordinate in the undeformed configuration, $\mu(R)$ ($=r(R)/R$) is the average azimuthal stretch with $r(R)$ denoting the average of the current radial coordinate along the thickness direction,  $\rho=r'(R)=\mu+R\mu'$ and $\lambda=\rho^{-1}\mu^{-1}$ are the radial and thickness stretches, respectively, $G(\rho,\mu)$ is such that it would be the free energy density of the homogeneous state if
$\rho$ and $\mu$ were constant, $G_1={\partial G}/{\partial\rho}$, and $P$ is the applied tension at the edge. The energy functional \eqref{eq:1d-introd} can be interpreted as follows:  the first term $\int_{0}^A G(\rho,\mu)R\,dR$ describes the total strain energy when gradient effects are neglected, which determines the amplitudes of thick and thin \lq\lq phases" in a ``two-phase" deformation; the other part of the integral in \rr{eq:1d-introd} accounts for the energy of the inhomogeneous interface, which describes how the two phases are connected. The Euler--Lagrange equation related to \eqref{eq:1d-introd} is a fourth-order nonlinear ordinary differential for $\mu(R)$, which is a significant simplification of the original nonlinear partial differential equations. The 1d model is validated by comparing its predictions for axisymmetric necking with finite-element simulations, which shows that the 1d model captures the entire evolution process of axisymmetric necking with surprising accuracy. The evolution of axisymmetric necking and the fate of the dielectric membrane subjected to mechanical and electric loads are investigated in detail  with the aid of this 1d model.

The rest of the paper is organized as follows. After summarizing in Section \ref{sec:2} the 3d nonlinear electroelasticity theory for a circular dielectric membrane,  we analyze in Section \ref{sec:3} the homogeneous deformation of the dielectric membrane and present the conditions for the pull-in instability and Treloar-Kearsley instability, and the bifurcation condition for axisymmetric necking. In Section \ref{sec:4}, we perform the dimension reduction and derive the aforementioned 1d model. Validation of the 1d model is conducted in Section \ref{sec:5}. In Section \ref{sec:6}, we apply the 1d model to investigate the evolution of axisymmetric necking in both weakly and fully nonlinear regimes. Finally, concluding remarks are given in Section \ref{sec:7}.

\section{Three-dimensional nonlinear electroelastic model}\label{sec:2}

\begin{figure}[h!]
	\centering
	\includegraphics[width=0.8\linewidth]{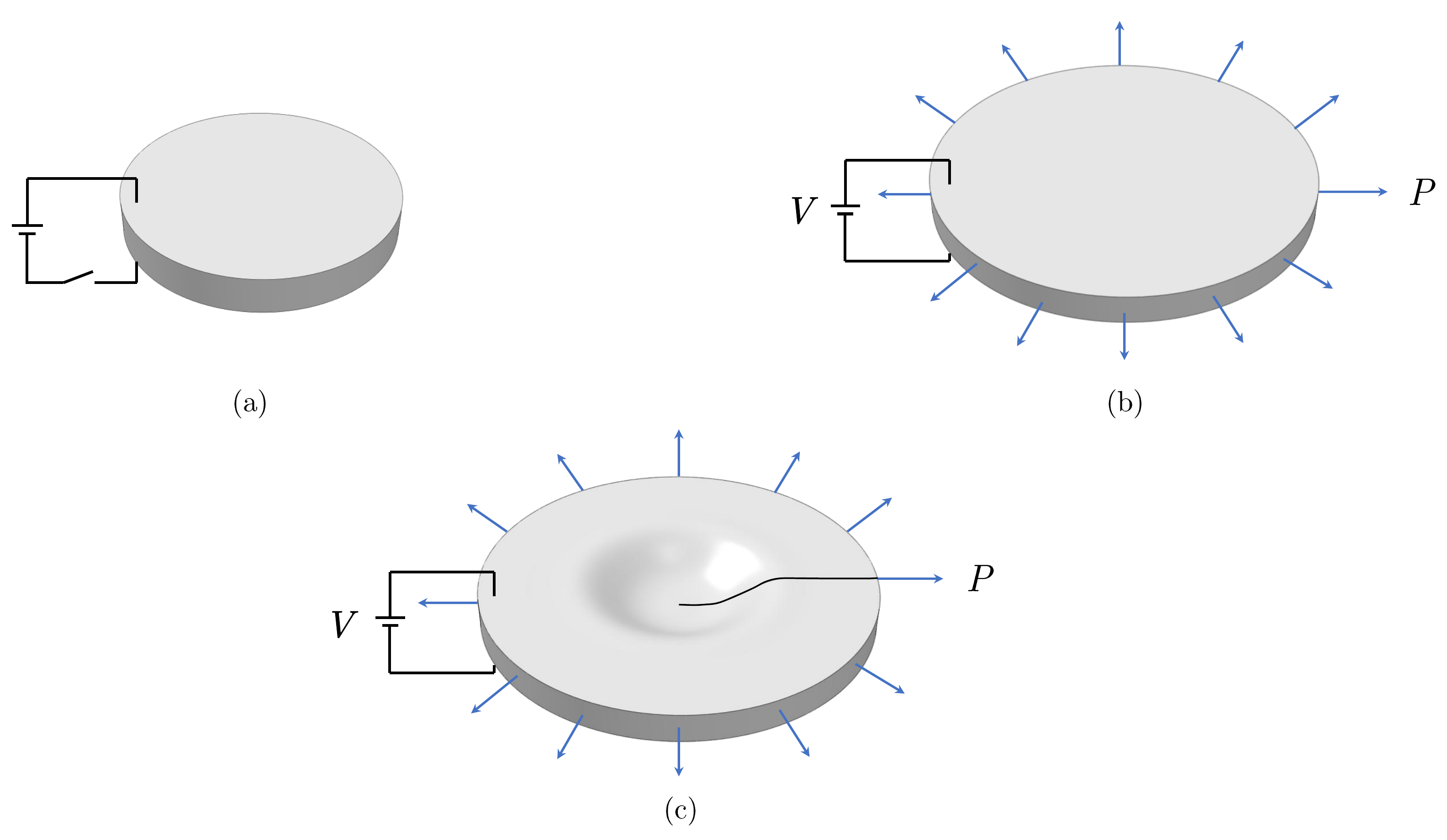}
	\caption{Three configurations of a circular dielectric plate subjected to  an  all-round  tension and an electric field: (a) reference (undeformed) configuration, (b) homogeneously deformed configuration and (c) current configuration.}
	\label{fig:plate}
\end{figure}

We consider an electrodes-coated dielectric plate that is subjected to equibiaxial stretching and an applied voltage between its upper and lower surfaces.
By focusing on a representative circular region of the dielectric plate, we may simply assume that the dielectric plate is circular with radius $A$ and thickness $H$, and an all-round tension is applied on its edge; see Fig.~\ref{fig:plate}(a).  The plate is thin so that the thickness-to-radius ratio $\varepsilon=H/A$ is small and we also refer to the plate as a membrane. Due to the combined effects of the
all-round tension and applied voltage, the dielectric membrane undergoes an axisymmetric deformation, resulting in a configuration that can be either homogeneous  or non-homogeneous, as shown in Fig.~ \ref{fig:plate}(b, c).

Cylindrical coordinates $(R,\Theta,Z)$ and $(r,\theta,z)$ are employed in the reference and current configurations, respectively. The common set of standard basis for both the reference and current coordinates is denoted by $(\bm{e}_r,\bm{e}_\theta,\bm{e}_z)$. The current position of a material point with reference cylindrical coordinates $(R,\Theta,Z)$ is then written as
\begin{align}
\bm{x}=r(R,Z)\bm{e}_r+z(R,Z)\bm{e}_z.
\end{align}
The deformation gradient is calculated as
\begin{align}\label{eq:F}
\bm{F}=r_R\bm{e}_r\otimes\bm{e}_r+r_Z\bm{e}_r\otimes\bm{e}_z+\frac{r}{R}\bm{e}_\theta\otimes\bm{e}_\theta+z_R\bm{e}_z\otimes\bm{e}_r+z_Z\bm{e}_z\otimes\bm{e}_z,
\end{align}
where the subscript indicates partial differentiation. The nominal electric field can be expressed in terms of a scalar electric potential $\varPhi(R,Z)$:
\begin{align}
\bm{E}=-\varPhi_R\bm{e}_r-\varPhi_Z\bm{e}_z.
\end{align}

In the framework of nonlinear electroelasticity \citep{dorfmann2014nonlinear},  the constitutive behavior of the dielectric plate is governed by a total free energy function $\Omega(\bm{F},\bm{E})$. We assume that the dielectric material is incompressible, isotropic with constant permittivity $\epsilon$ (the so-called ideal dielectric), corresponding to the total free energy of the form
\begin{align}\label{eq:Omega}
\Omega(\bm{F},\bm{E})=W(\lambda_1,\lambda_2,\lambda_3)-\frac{1}{2}\epsilon \bm{E}\cdot\bm{C}^{-1}\bm{E},
\end{align}
where $W$ is the elastic strain energy function, $\lambda_1$, $\lambda_2$ and $\lambda_3$ are the principal stretches and $\bm{C}=\bm{F}^T\bm{F}$ is the right Cauchy-Green deformation tensor.  Here and hereafter, we identify the indices $1$, $2$, $3$ such that for homogeneous deformations they coincide with the $r$-, $\theta$-, and $z$-directions, respectively. The permittivity $\epsilon$ is recognized to be deformation-dependent in general \citep{jimenez2013deformation, cohen2016towards,jimenez2016constitutive,dorfmann2018effect}. Our methodology could incorporate this extra complexity without any
technical difficulties, but we adopt the ideal dielectric assumption in order to simplify our presentation.

We use a variational formulation to keep the problem description as simple as possible. The total potential energy of the plate is composed of the total free energy  and load potential due to the all-round tension, which reads, after scaling by $2 \pi$,
\begin{align}\label{eq:E}
\begin{split}
\mathcal{E}[r,z,\varPhi]=\int_0^A \int_{-H/2}^{H/2} \Big(w(\lambda_1,\lambda_2)-\frac{1}{2}\epsilon \bm{E}\cdot\bm{C}^{-1}\bm{E}\Big)\,dZ R\, dR-\int_{-H/2}^{H/2} P A r(A,Z)\,dZ,
\end{split}
\end{align}
where $w(\lambda_1,\lambda_2)=W(\lambda_1,\lambda_2,\lambda_1^{-1}\lambda_2^{-1})$ is the reduced strain energy function, and $P$ is the applied tensile force (per unit undeformed area) on the edge surface. The upper and lower surfaces are assumed to be traction-free, but there the electric potential $\varPhi$ is specified according to
\begin{align}\label{eq:phi}
\varPhi(R,-\frac{H}{2})=-\frac{V}{2},\quad \varPhi(R,\frac{H}{2})=\frac{V}{2}
\end{align}
with $V$ being the applied voltage.  The nonlinear electroelastic model governed by the energy functional \eqref{eq:E} will be used as the starting point for the subsequent dimension reduction.

\section{Homogeneous deformation}\label{sec:3}

Let us first summarize the main results concerning the homogeneous deformation of a dielectric plate under biaxial stretching. In Cartesian coordinates,  a homogeneous biaxial deformation takes  the form
\begin{align}
x_1=\lambda_1 X_1,\quad x_2=\lambda_2 X_2,\quad z=\lambda_3 Z,\quad E_3=-\frac{V}{H},
\end{align}
where $(X_1,X_2,Z)$ and $(x_1,x_2,z)$ are the reference and current Cartesian coordinates, $\lambda_1$, $\lambda_2$ and $\lambda_3$ are the (constant) in-plane and thickness stretches, respectively, and $E_3$ is the nominal electric field in the thickness direction. Note that here and in this section only, the indices $1$, $2$  refer to the $X_1$- and $X_2$-directions of Cartesian coordinates, respectively.

 The incompressibility constraint then implies  $\lambda_3=\lambda_1^{-1}\lambda_2^{-1}$, so that the total free energy function  reduces to
\begin{align}
\Omega(\bm{F},\bm{E})=\tilde{\Omega}(\lambda_1,\lambda_2,E_3):=w(\lambda_1,\lambda_2)-\frac{1}{2}\epsilon \lambda_1^2\lambda_2^2 E_3^2.
\end{align}
Specializing to the equibiaxial  case where $\lambda_1=\lambda_2=\mu$, we see that the total potential energy \eqref{eq:E} is given by
\begin{align}
\mathcal{E}_\text{hom}=\frac{1}{2}A^2 H\Big(w(\mu,\mu)-\frac{1}{2}  \mu^4 U^2\Big) -P A^2H \mu,
\end{align}
where {$U=\sqrt{\epsilon}V/H$} is the scaled voltage. From the equilibrium equation $ d\mathcal{E}_\text{hom}/d\mu=0$, we obtain
\begin{align}\label{eq:P}
P=\frac{1}{2}\frac{d}{d\mu}w(\mu,\mu)-\mu^3 U^2.
\end{align}
The above equation gives the relation between the load parameters $P$ and $U$ and the deformation parameter $\mu$ for the  homogeneous equibiaxial deformation.

Associated with the stretches $\lambda_1$ and $\lambda_2$ and nominal electric field $E_3$, the nominal stresses $S_1$ and $S_2$ and nominal electric displacement $D_3$ are given by \citep{dorfmann2005nonlinear}
\begin{align}
S_1=\frac{\partial\tilde{\Omega}}{\partial\lambda_1},\quad S_2=\frac{\partial\tilde{\Omega}}{\partial\lambda_2},\quad D_3=-\frac{\partial\tilde{\Omega}}{\partial E_3}.
\end{align}
In terms of the Legendre transform $\Omega^*(\lambda_1,\lambda_2,D_3)=\tilde{\Omega}(\lambda_1,\lambda_2,E_3)+E_3D_3$, the above relation can also be written as
\begin{align}
S_1=\frac{\partial\Omega^*}{\partial\lambda_1},\quad S_2=\frac{\partial\Omega^*}{\partial\lambda_2},\quad E_3=\frac{\partial\Omega^*}{\partial D_3}.
\end{align}
The Hessian stability criterion \citep{zhao2007method} asserts that the Hessian matrix of the function $\Omega^*(\lambda_1,\lambda_2,D_3)$ defined by
\begin{align}
\bm{H}_{\Omega^*}=\begin{pmatrix}
\frac{\partial^2\Omega^*}{\partial\lambda_1^2} & \tfrac{\partial^2\Omega^*}{\partial\lambda_1\partial\lambda_2} & \frac{\partial^2\Omega^*}{\partial\lambda_1\partial D_3}\\
\frac{\partial^2\Omega^*}{\partial\lambda_2\partial\lambda_1} & \frac{\partial^2\Omega^*}{\partial\lambda_2^2} & \frac{\partial^2\Omega^*}{\partial\lambda_2\partial D_3}\\
\frac{\partial^2\Omega^*}{\partial D_3 \partial\lambda_1} & \frac{\partial^2\Omega^*}{\partial D_3 \partial\lambda_2} & \frac{\partial^2\Omega^*}{\partial D_3^2}
\end{pmatrix}=\begin{pmatrix}
\frac{\partial S_1}{\partial\lambda_1} & \frac{\partial S_1}{\partial\lambda_2} & \frac{\partial S_1}{\partial D_3}\\
\frac{\partial S_2}{\partial\lambda_1} & \frac{\partial S_2}{\partial\lambda_2} & \frac{\partial S_2}{\partial D_3}\\
\frac{\partial E_3}{\partial\lambda_1} & \frac{\partial E_3}{\partial\lambda_2} & \frac{\partial E_3}{\partial D_3}
\end{pmatrix}
\end{align}
should be positive definite to ensure stability.

In particular, when the determinant of the Hessian matrix vanishes, an instability may occur. Evaluating the determinant of the Hessian matrix at the equibiaxial stretching $\lambda_1=\lambda_2=\mu$ where $S_1=S_2=S(\mu,D_3)$, $E_3=E(\mu,D_3)$, ${\partial S_1}/{\partial\lambda_1}={\partial S_2}/{\partial\lambda_2}$, ${\partial S_1}/{\partial\lambda_2}={\partial S_2}/{\partial\lambda_1}$ and ${\partial E_3}/{\partial\lambda_1}={\partial E_3}/{\partial\lambda_2}$, we obtain
\begin{align}\label{eq:HD}
\det(\bm{H}_{\Omega^*})=\Big(\frac{\partial S_1}{\partial\lambda_1}-\frac{\partial S_1}{\partial\lambda_2}\Big)\Big(\frac{\partial S}{\partial \mu}\frac{\partial E}{\partial D_3}-\frac{\partial S}{\partial D_3}\frac{\partial E}{\partial \mu}\Big),
\end{align}
where all quantities are evaluated at $\lambda_1=\lambda_2=\mu$.

It follows from \eqref{eq:HD} that $\det(\bm{H}_{\Omega^*})=0$ is satisfied if one of the following two conditions is satisfied:
\begin{align}
&\frac{\partial S_1}{\partial\lambda_1}-\frac{\partial S_1}{\partial\lambda_2}=\lim_{\lambda_2\to\lambda_1}\frac{S_2-S_1}{\lambda_2-\lambda_1}=0, \la{3.9t}\\
&\frac{\partial S}{\partial\mu}\frac{\partial E}{\partial D_3}-\frac{\partial S}{\partial D_3}\frac{\partial E}{\partial \mu}=0.\la{3.10t}
\end{align}
It is clear that the first equation is the bifurcation condition for the Treloar-Kearsley instability where unequal stretches occur at equal nominal stresses \citep{ogden1985local, kearsley1986asymmetric,ogden1987stability}. Also, it can be shown that the second equation is equivalent to
\begin{align}
\frac{d E}{d D_3}\Big|_{S\ \text{fixed}}=0,\quad \frac{d S}{d \mu}\Big|_{E\ \text{fixed} }=0,
\end{align}
so it signifies the onset of pull-in instability \citep{zhao2007method,norris2008comment,chen2021interplay}. In contrast, it was shown in \cite{fu2023axisymmetric} that the bifurcation condition for localized axisymmetric necking is given by
\begin{align}\label{added11}
\frac{\partial S_1}{\partial\lambda_1}\Big|_{\lambda_1=\lambda_2=\mu} =0,
\end{align}
which is different from both \rr{3.9t} and \rr{3.10t}. This indicates that the Hessian stability criterion is insufficient to determine all instability modes of dielectric elastomers. For a discussion of how the three instabilities compete for different material models, we refer to \cite{fu2023axisymmetric}.

\section{Derivation of the one-dimensional  model} \label{sec:4}

Subjected to equibiaxial tension and an electric field, a circular dielectric plate may suffer from axisymmetric necking that initiates at the center, then grows in amplitude (corresponding to an increased reduction in thickness), and finally propagates towards the edge \citep{fu2023axisymmetric}. To describe this entire evolution process semi-analytically, we derive in this section a 1d reduced model from the 3d electroelasticity using the variational asymptotic method.

\subsection{Expansion of the energy functional}

The starting point of our derivation is the observation that the necking deformation varies slowly in the radial direction, at least in its early stage, due to its long wavelength. This motivates us to define  a ``far-distance" variable $S$ by
\begin{align}
S=\varepsilon R,
\end{align}
where  $\varepsilon=H/A\ll 1$ is the small thickness-to-radius ratio. The slowly varying property implies that the dependence of all state variables on $R$ is through $S$. Guided by the scalings in previous work \citep{audoly2016analysis,yu2023one}, we look for an asymptotic solution of the form
\begin{align}\label{eq:rz}
\begin{split}
&r(R,Z)=\varepsilon^{-1}\mu(S)S+\varepsilon u^*(S,Z)+O(\varepsilon^3),\\
&z(R,Z)=\lambda(S)Z+\varepsilon^2 v^*(S,Z)+O(\varepsilon^4),\\
&\varPhi(R,Z)=\frac{V}{H} Z+\varepsilon^2 \varphi^*(S,Z)+O(\varepsilon^4),
\end{split}
\end{align}
where $\mu(S)$ is the leading-order azimuthal stretch, $\lambda(S)$ is chosen to satisfy the incompressibility constraint at leading order:
\begin{align}
	\lambda(S)=\mu(S)^{-1}(\mu(S)+S\mu'(S))^{-1}, \la{lambda}
\end{align}
and $u^*$ is assumed to satisfy the kinematic constraint
\begin{align}\label{eq:ucon}
\int_{-H/2}^{H/2} u^*(S,Z)\,dZ=0.
\end{align}
In view of \eqref{eq:phi}, $\varphi^*$ satisfies the boundary conditions
\begin{align}\label{eq:varphi}
\varphi^*(S,-\frac{H}{2})=0,\quad \varphi^*(S,\frac{H}{2})=0.
\end{align}
Note that the leading-order terms of \eqref{eq:rz} are modified from the homogeneous solution  by allowing the stretches to be functions of $S$ so as to describe non-homogeneous deformations; the  next-order terms  are higher-order corrections that are added to restore the self-consistency of the expansion. Also, the consequence of  \eqref{eq:ucon} is that $\mu(S)$ represents the average of the azimuthal stretch along the thickness direction, which helps to simplify the dimension reduction.  In the work of \cite{zurlo2017catastrophic} and \cite{greaney2019out}, the authors used an approximation that includes only the leading-order terms. This approximation is not asymptotically consistent, and is equivalent to setting the correction terms to zero in our approach, which is not optimal in view of \eqref{eq:u} to be derived later. As a result, the leading-order approximation leads to a reduced model  with an overestimated gradient coefficient \citep{audoly2016analysis}.

We use \eqref{eq:rz} to calculate the deformation gradient $\bm{F}$ and nominal electric field $\bm{E}$ to order $\varepsilon^2$, obtaining
\begin{align}
&\bm{F}=
\begin{pmatrix}
\rho(S)+\varepsilon^2 u_S^* & 0 & \varepsilon u^*_Z\\
0 & \mu(S)+\varepsilon^2 u^*/S & 0\\
\varepsilon\lambda'(S)Z & 0 & \lambda(S)+\varepsilon^2 v_Z^*
\end{pmatrix}, \label{eq:FF}\\
&\bm{E}=(0,0,-\epsilon^{-1/2}U-\varepsilon^2\varphi^*_Z), \label{eq:e}
\end{align}
where $\rho(S)=\mu(S)+S\mu'(S)$ denotes the leading-order radial stretch. The incompressibility constraint $\det(\bm{F})=1$ requires
\begin{align}\label{eq:incom}
\rho\lambda u^*+ S\mu (\rho v^*_Z+\lambda u^*_S-Z\lambda' u^*_Z)=0.
\end{align}
It follows from \eqref{eq:FF} that the principal stretches (eigenvalues of  $\sqrt{\bm{F}^T\bm{F}}$)  are
\begin{align}
\begin{split}\label{eq:lam}
\lambda_1&=\rho+\varepsilon^2 \Big(\frac{\rho(u^{*2}_Z+\lambda'^2 Z^2)+2\lambda\lambda' Z  u^*_Z}{2(\rho^2-\lambda^2)}+u^*_S\Big)+O(\ep^3),\\
\lambda_2&=\mu+\varepsilon^2 \frac{u^*}{S}+O(\ep^3),
\end{split}
\end{align}
where the expression for $\lambda_3$ is omitted as it is not required in the subsequent derivation.

Based on \eqref{eq:FF}, \eqref{eq:e} and \eqref{eq:lam}, we can expand the energy functional \eqref{eq:E}  as
\begin{align}\label{eq:EE}
\mathcal{E}[\mu,u]=H \Big(\int_0^A  (w(\rho,\mu)-\frac{1}{2} \rho^2\mu^2 U^2 )R\,dR-PA^2 \mu (A)\Big)+\mathcal{E}_2+O(A^2\varepsilon^3),
\end{align}
where $\mathcal{E}_2$ represents the term of order $\varepsilon^2$ and can be written, in terms of the un-scaled variables, as
\begin{align}\label{eq:EE2}
\begin{split}
\mathcal{E}_2=&\int_0^A  \int_{-H/2}^{H/2}  \Big(w_1\frac{\rho (u^{2}_Z+\lambda'^2 Z^2)+2 \lambda\lambda' Z  u_Z}{2(\rho^2-\lambda^2)}-\frac{1}{2}\mu^2 U^2\lambda'^2 Z^2 \Big)\,dZ R\,dR.
\end{split}
\end{align}
In the above expression, $u(R,Z)=\varepsilon u^*(S,Z)$  denotes the un-scaled displacement, we have written $\mu(R)$, $\rho(R)$,  $\lambda(R)$ for $\mu(S)$, $\rho(S)$, $\lambda(S)$ respectively to avoid introducing extra notation  so that $\lambda'$ now stands for $\lambda'(R)$, and $w_1=\frac{\partial w}{\partial\rho}(\rho,\mu)$ in which $\rho$ and $\mu$ are now functions of $R$. Note that use has also been made of equations \eqref{eq:ucon}, $\eqref{eq:varphi}$ and \eqref{eq:incom} to reduce $\mathcal{E}_2$ to the form \eqref{eq:EE2}.

\subsection{Optimal correction}
By the principle of stationary potential energy, we need to find $\mu$ and $u$ such that \eqref{eq:EE} is stationary. According to the variational asymptotic method \citep{berdichevskii1979variational,audoly2016analysis,lestringant2020asymptotically}, this task can be made simpler by using a two-step variational procedure. First, we treat $\mu$ as stipulated and seek $u$ such that $\eqref{eq:EE}$ is stationary.
In view of the facts that
\begin{align}
\rho (u^{2}_Z+\lambda'^2 Z^2)+2 \lambda\lambda' Z  u_Z=\rho\Big(u_Z+\frac{\lambda\lambda'}{\rho}Z\Big)^2+\frac{\rho^2-\lambda^2}{\rho}\lambda'^2Z^2,
\end{align}
and that $\rho$, $\lambda$ and $w_1$ are all independent of $Z$, we conclude that $\mathcal{E}_2$, and consequently \eqref{eq:EE}, are minimized when $u_Z=-\frac{\lambda\lambda'}{\rho}Z$. Integrating this equation subject to \eqref{eq:ucon} then yields the optimal correction as
\begin{align}\label{eq:u}
u=-\frac{\lambda\lambda'}{2\rho}\Big(Z^2-\frac{1}{12}H^2\Big).
\end{align}

An important feature of the above two-step variational method is that it is equivalent to the original simultaneous variation. 
To illustrate this, consider the problem of minimizing the function $f(x,y)$. According to the two-step minimization procedure, we first treat $x$ as prescribed  and seek $y$ such that $f(x,y)$ is minimal, which implies the optimal $y=y_\text{opt}(x)$ satisfies $f_y(x,y_\text{opt}(x))=0$, where the subscript $y$ signifies partial differentiation. Inserting this into $f(x,y)$ and minimizing the resulting function $g(x):=f(x,y_\text{opt}(x))$, we obtain
	\begin{align}\label{eq:g}
	g'(x)=f_x(x,y_\text{opt}(x))+f_y(x,y_\text{opt}(x))y'_\text{opt}(x)=f_x(x,y_\text{opt}(x))=0.
	\end{align}
The solution $x_0$ to \eqref{eq:g}  satisfies $f_x(x_0,y_\text{opt}(x_0))=f_y(x_0,y_\text{opt}(x_0))=0$, justifying the equivalence between the two-step minimization and  simultaneous minimization.  \cite{audoly2021asymptotic} verified this equivalence for functionals.

\subsection{One-dimensional energy functional}

Inserting the optimal solution \eqref{eq:u} back into \eqref{eq:EE} and simplifying, followed by dividing the expression by $H$,  we  obtain the following 1d energy functional that captures the gradient effect:
\begin{align}\label{eq:1d}
\begin{split}
\mathcal{E}_{\text{1d}}[\mu]&=\int_0^A  \Big(G(\rho,\mu) +\frac{H^2}{24}\frac{  G_1(\rho,\mu)}{\rho}\lambda'(R)^2\Big)R- PA^2\mu(A),
\end{split}
\end{align}
where $\rho=\mu+R\mu'$ and $\lambda=\mu^{-1}(\mu+R\mu')^{-1}$ are the leading-order radial and thickness stretches, respectively, and the moduli $G$ and $G_1$ are given by
\begin{align}\label{eq:GG}
G(\rho,\mu)=w(\rho,\mu)-\frac{1}{2} \rho^2\mu^2 U^2,\quad G_1(\rho,\mu)=\frac{\partial G(\rho,\mu)}{\partial\rho}=w_1(\rho,\mu)- \rho\mu^2 U^2.
\end{align}
Note that the stretches $\mu$, $\rho$ and $\lambda$ are functions of $R$, with the independent variable  left out for readability. The second term in \eqref{eq:1d} is a strain-gradient term accommodating non-homogeneous deformations. The 1d energy functional \eqref{eq:1d} has the form that  involves the derivatives of $\mu$ to second order:
\begin{align}
&\mathcal{E}_{\text{1d}}[\mu]=\int_0^A L(R, \mu, \mu', \mu'')\,dR-PA^2\mu(A),
\end{align}
where the integrand $L$ is given by
\begin{align}
&L(R, \mu, \mu', \mu'')=\Big(G(\rho,\mu) +\frac{H^2}{24}\frac{  G_1(\rho,\mu)}{\rho^5\mu^4}(3\mu\mu'+R\mu'^2+R\mu\mu'')^2\Big)R.
\end{align}

  Finally, extremizing \eqref{eq:1d} with respect to $\mu$ yields the Euler--Lagrange equation of equilibrium and the associated boundary conditions:
 \begin{align}
 &\frac{\partial L}{\partial \mu}-\frac{d}{dR}\Big(\frac{\partial L}{\partial \mu'}\Big)+\frac{d^2}{dR^2}\Big(\frac{\partial L}{\partial \mu''}\Big)=0,\label{eq:EL}\\
 &\frac{\partial L}{\partial \mu'}-\frac{d}{dR}\Big(\frac{\partial L}{\partial \mu''}\Big)=PA^2\quad  \text{at}\ R=A, \label{eq:bc1}\\
 &3\mu\mu'+R\mu'^2+R\mu\mu''=0\quad\text{at}\ R=A. \label{eq:bc2}
 \end{align}
We remark that  \eqref{eq:bc2} corresponds to the natural boundary conditions  $\partial L/\partial\mu''=0$ at $R=A$ and can also be written as $\lambda'(A)=0$.  In the displacement-controlled problems,  \eqref{eq:bc1} should be replaced by $\mu(A)=\hat{a}$ with $\hat{a}$ being a prescribed constant. The above boundary value problem is to be solved subject to the regularity conditions
\be
 \mu'(0)=0, \quad \mu'''(0)=0. \label{eq:bc0}\en
This boundary value problem is our 1d  reduced  model that governs the variation of the current radius of the plate (which is $R$ times $\mu(R)$). Once $\mu(R)$ is determined, the necking profile (i.e., thickness variation), described by $\lambda(R)$, can be obtained with the aid of \rr{lambda}.

\section{Validation of the one-dimensional model}\label{sec:5}

In this section, we validate the 1d model by comparing its predictions for axisymmetric necking with those of the exact 3d theory and Abaqus simulations. We shall show that the 1d model can produce the same weakly nonlinear necking solution as that based on the exact 3d theory in the near-critical regime and yields sufficiently accurate  solutions in line with Abaqus simulations in the entire post-bifurcation regime.

To minimize the algebra, we conduct the validation for the purely mechanical case when the plate is only subjected to an all-round tension, that is, the electric load $U=0$. Also, to avoid ambiguity, we use $a$ as a substitute for $\mu$  in the case of homogeneous deformation and set $a=\mu(A)$ if the deformation is nonhomogeneous. With this convention, the axisymmetric homogeneous solution is written as
\begin{align}
r=a R,\quad z=a^{-2}Z,
\end{align}
where  $a$ is a constant. It follows from \eqref{eq:P} that $a$ satisfies $P=\frac{1}{2}\frac{d}{da}w(a,a)$.

\subsection{Comparison with the exact weakly nonlinear analysis}\label{sec:linear}
To study the bifurcation of the above homogeneous solution, we consider a perturbed solution of the form
\begin{align}\label{eq:a}
\mu(R)=a+y(R),
\end{align}
where $y(R)$ is a small perturbation. Substituting \eqref{eq:a} into the 1d equilibrium equations \eqref{eq:EL} with $U=0$ and ignoring nonlinear terms, we obtain
\begin{align}
&y^{(4)}(R)+\frac{6}{R}y^{(3)}(R)+\Big(\omega(a)+\frac{3}{R^2}\Big)y''(R)+\frac{3}{R}\Big(\omega(a)-\frac{1}{R^2}\Big)y'(R)=0, \label{eq:lin1}
\end{align}
where
\be \omega(a)=-\frac{12a^7 w_{11}(a,a)}{H^2 w_1(a,a)}, \;\;\;\;  w_{11}(\rho, \mu)= \frac{\partial^2 w}{\partial \rho^2}. \la{oct1} \en The general solution to \eqref{eq:lin1} bounded at $R=0$ is given by
\begin{align}\label{eq:ylinear}
y(R)=b_1+b_2 \frac{J_1(\sqrt{\omega(a)}R)}{R},
\end{align}
where $b_1$ and $b_2$ are arbitrary constants, and $J_1(x)$ is the Bessel function of the first kind. The function $J_1(\sqrt{\omega(a)}R)$ is
oscillatory with $\sqrt{\omega(a)}$ playing the role of a wavenumber. According to the analysis in \cite{wang2022axisymmetric}, we anticipate that axisymmetric necking occurs  when this wavenumber vanishes: $\sqrt{\omega(a)}=0$, which is equivalent to
\begin{align}\label{eq:bif}
w_{11}(a,a)=0.
\end{align}
This recovers the necking condition derived by \cite{wang2022axisymmetric} from the 3d theory.

Next, we  conduct a weakly nonlinear analysis to derive the amplitude equation for the axisymmetric necking. We use $a=\mu(A)$ as the control parameter in our near-critical analysis, and denote the critical stretch, the solution to \eqref{eq:bif}, by $a_\text{cr}$.  Guided by the scalings in \cite{wang2022axisymmetric}, we may write
\begin{align}
a=a_\text{cr}+\eta a_0,
\end{align}
where $a_0$ is a constant and $\eta$ is a positive small parameter. It is seen from \eqref{eq:ylinear} that $y$ depends on $R$ through the ``far-distance" variable $X=\sqrt{\eta}R$ in the near-critical regime. Thus, we look for an asymptotic solution of the form
\begin{align}\label{eq:yy}
\mu(R)=a+\sqrt{\eta} \{ y_1(X)+\eta y_2(X)+\cdots \},
\end{align}
where the functions $y_1(X)$, $y_2(X)$, $\dots$ are to be determined. On substituting \eqref{eq:yy} into  \eqref{eq:EL} with $U=0$ and equating the coefficients of like powers of $\eta$, we obtain a hierarchy of equations. The leading-order equation reproduces the bifurcation condition  \eqref{eq:bif}. At the next order, we obtain
\begin{align}\label{eq:weakly}
\begin{split}
&y_1^{(4)}(X)+\frac{6}{X} y_1^{(3)}(X) + \frac{3}{X^2} y_1''(X)-\frac{3}{X^3}y_1'(X) +c_1 a_0 \Big(y_1''(X)+ \frac{3}{X}y_1'(X) \Big)\\
&+c_1\Big(y_1''(X) y_1(X)+\frac{3}{X} y_1'(X)y_1(X) \Big)+c_2 X y_1'(X)y_1''(X)+c_3 y_1'(X)^2=0,
\end{split}
\end{align}
where the constant coefficients $c_1$, $c_2$ and $c_3$ are given by
\begin{align}\label{eq:wamp}
\begin{split}
&c_1=-\frac{12 a_\text{cr}^7}{H^2} \frac{w_{111}(a_\text{cr},a_\text{cr})+w_{112}(a_\text{cr},a_\text{cr})}{w_1(a_\text{cr},a_\text{cr})},\\
&c_2=-\frac{12 a_\text{cr}^7}{H^2}\frac{w_{111}(a_\text{cr},a_\text{cr})}{w_1(a_\text{cr},a_\text{cr})},\\
&c_3=-\frac{6 a_\text{cr}^7}{H^2}\frac{5w_{111}(a_\text{cr},a_\text{cr})+w_{112}(a_\text{cr},a_\text{cr})}{w_1(a_\text{cr},a_\text{cr})}.
\end{split}
\end{align}
 In the above expressions, the two higher order moduli $w_{111}$ and $w_{112}$ are defined by
 \be
 w_{111}(\rho, \mu)=\frac{\partial^3 w}{\partial\rho^3}, \;\;\;\; w_{112}(\rho, \mu)=\frac{\partial^3 w}{\partial\rho^2\partial\mu}. \la{oct3} \en
 Note that the coefficients satisfy the relation $c_1+4c_2-2c_3=0$, since they involve only two third-order derivatives of $w$.
  One can check that the amplitude equation \eqref{eq:weakly} is identical to its counterpart Eq.~(52) in \cite{wang2022axisymmetric} which was derived using the exact 3d theory. Note, however, that our quantities $(a_\text{cr},H,y_1(X))$ here correspond to their $(\lambda_\text{cr},\lambda_\text{cr}^2h,{A(\lambda_\text{cr} X)}/{X})$.

The fourth-order differential equation \eqref{eq:weakly} subject to $y'(0)=0$, $y'''(0)=0$ and appropriate edge conditions can be solved using the standard finite difference method. We refer to \cite{fu2023axisymmetric} for further details.

\subsection{Comparison with Abaqus simulations}

To compare the predictions of the 1d model with full Abaqus simulations, we need to solve the 1d model in the fully nonlinear regime.
The Rayleigh-Ritz method \citep{dhatt2012finite} is used, which is a  direct numerical method for minimizing a given functional. It is direct in the sense that it yields a solution to the variational problem without solving the associated Euler-Lagrange equation, making it well suited for the present problem considering the complex expression of the Euler--Lagrange equation. It should be mentioned that the Rayleigh-Ritz method has been shown to be equivalent to the lattice method \citep{yashin2007theoretical,zhang2019deriving}; the former is commonly used  in the numerical mathematics community, while the latter is preferred by the mechanics community.

In the Rayleigh-Ritz method, we discretize the 1d functional \eqref{eq:1d} using an approximation of $\mu$ obtained by interpolation and then obtain its stationary conditions with respect to the discrete variables. The presence of $\mu''$ in \eqref{eq:1d} requires the use of cubic Hermite interpolation to ensure the continuity of $\mu'$, but the use of such a higher-order polynomial interpolation would make the numerical scheme more complex and less efficient. To resolve this issue, we view $\lambda$ as an independent variable satisfying the incompressibility constraint $\mu+R\mu'-\mu^{-1}\lambda^{-1}=0$. By the method of Lagrange multipliers, minimizing \eqref{eq:1d} with $U=0$ is equivalent to minimizing the Lagrangian functional
\begin{align}\label{eq:E1dL}
\tilde{\mathcal{E}}_\text{1d}[\mu,\lambda,q]=\int_0^A \Big(w(\rho,\mu) +\frac{H^2}{24}\frac{  w_1(\rho,\mu)}{\rho}\lambda'(R)^2-q(\mu+R\mu'-\rho)\Big)R\,dR-P A^2\mu(A),
\end{align}
where we have written $\rho$ for $\mu^{-1}\lambda^{-1}$, and $q$ is the Lagrange multiplier to enforce the constraint $\mu+R\mu'-\mu^{-1}\lambda^{-1}=0$.  Then we partition the domain $[0,A]$ into $n$ equal intervals and approximate the functions $\mu$, $\lambda$ and $q$ on each interval using linear interpolation in terms of their nodal values. For instance, on the $j$-th interval, $\mu(R)$ is approximated by the linear function
\begin{align}\label{eq:mu}
\mu(R)=\mu_j+\frac{\mu_{j+1}-\mu_j}{h}(R-jh),\quad jh\leq R\leq (j+1)h,
\end{align}
where $h=A/n$ is the mesh size. Substituting \eqref{eq:mu} and similar approximations for $\lambda$ and $q$ into \eqref{eq:E1dL} and calculating the integration in \eqref{eq:E1dL} on each interval using a two-point Gaussian quadrature rule, we obtain a discrete energy in terms of the nodal values:
\begin{align}\label{eq:E1dLL}
\tilde{\mathcal{E}}_\text{1d}[\mu,\lambda,q]=\hat{\mathcal{E}}_\text{1d}(\mu_j,\lambda_j,q_j),\quad j=0,1,\dots,n.
\end{align}
From the stationary conditions of \eqref{eq:E1dLL}, we obtain $3n+3$ equations:
\begin{align}\label{eq:diseq}
\begin{split}
&\frac{\partial\hat{\mathcal{E}}_\text{1d}}{\partial \mu_j}=0,\quad  \frac{\partial\hat{\mathcal{E}}_\text{1d}}{\partial \lambda_j}=0,\quad \frac{\partial\hat{\mathcal{E}}_\text{1d}}{\partial q_j}=0,\quad j=0,1,\dots,n.
\end{split}
\end{align}
In the displacement-controlled problems, ${\partial\hat{\mathcal{E}}_\text{1d}}/{\partial \mu_n}=0$ should be replaced by $\mu_n=\hat{a}$. The above system of algebraic equations can be solved using the Newton-Raphson method with a suitable initial guess. One may use the weakly nonlinear solution  as an initial guess for \eqref{eq:diseq}  in the near-critical region and continue the solution to the fully nonlinear regime by always using the solution at the previous step as the initial guess for the current step. The above numerical scheme is implemented in the symbolic computation software Mathematica \citep{wolfram2024mathematica}  and it is found that taking $n=1000$ yields sufficiently accurate results. The above system of equations  can typically be solved in a few seconds on a personal computer, which is much faster than Abaqus simulations.

 In  Abaqus simulations, the circular plate is simulated using an axisymmetric model and meshed by an eight-node reduced-integration hybrid quadratic axisymmetric element (Abaqus element CAX8RH). Also, to ensure that necking occurs at the center of the plate, a circular region with a radius of $0.1A$ around the center is weakened by taking its shear modulus  to be $0.9999$ times that of the rest of the plate.

For the purpose of illustration, we consider the following incompressible Ogden model \citep{plante2006large,wang2022axisymmetric}
\begin{align}\label{eq:mm}
w(\lambda_1,\lambda_2)=\frac{2\mu_1}{m_1^2}(\lambda_1^{m_1}+\lambda_2^{m_1}+\lambda_1^{-m_1}\lambda_2^{-m_1}-3)+\frac{2\mu_2}{m_2^2}(\lambda_1^{m_2}+\lambda_2^{m_2}+\lambda_1^{-m_2}\lambda_2^{-m_2}-3)
\end{align}
with $\mu_2/\mu_1=1/80$, $m_1=1/2$, $m_2=4$. The undeformed radius is set to be $A=5H$. By scaling all stress variables by $\mu_1$ and length variables by $H/2$, we may take $\mu_1=1$ and $H=2$. In particular, now $A=10$.  For the material model given above, it follows from \eqref{eq:bif} that the bifurcation occurs at $a=a_\text{cr}=2.439$ with a critical tension $P_\text{cr}=1.98$. As we trace the bifurcation solution away from the bifurcation point,  the radius of the plate decreases and the necking grows until it reaches a maximum amplitude. The  maximum of necking is attained at $a=a_M=2.172$ according to  the numerical continuation method. After that, the necking propagates radially towards the edge and the radius increases. We remark that we had to change the control parameter in the 1d numerical continuation  due to the existence of  turning points in the loading curve:  $\mu(0)$ is used as the control parameter in the growth stage of necking and $a$ is used as the control parameter in the propagation stage of necking.

\begin{figure}[h!]
	\centering
	\subfloat[]{\includegraphics[width=0.4\textwidth]{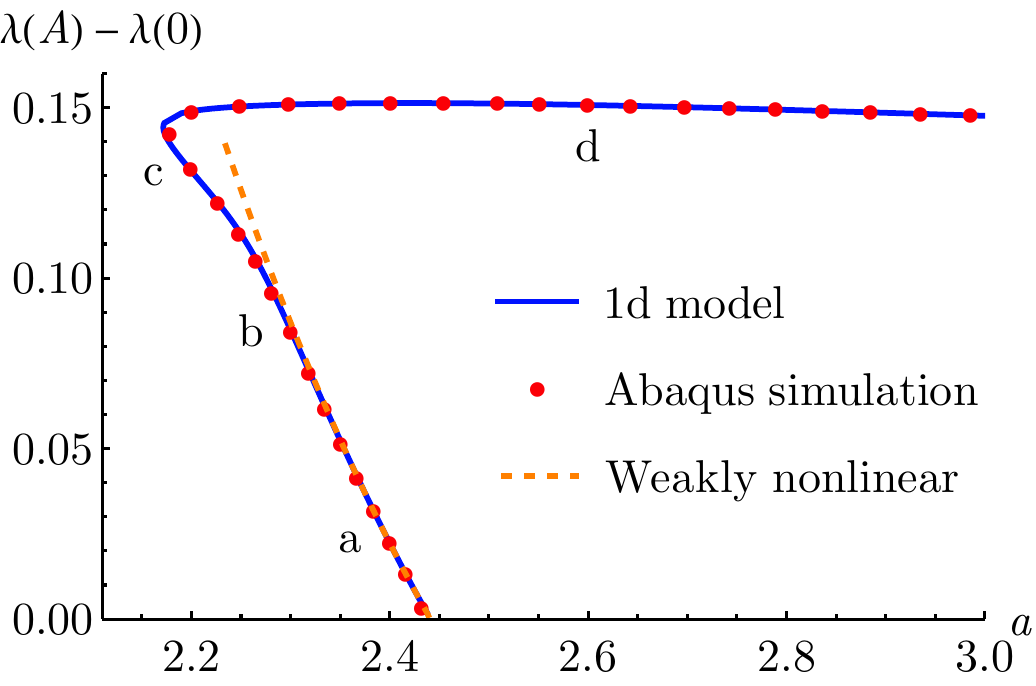}
	}\qquad\qquad
	\subfloat[]{\includegraphics[width=0.4\textwidth]{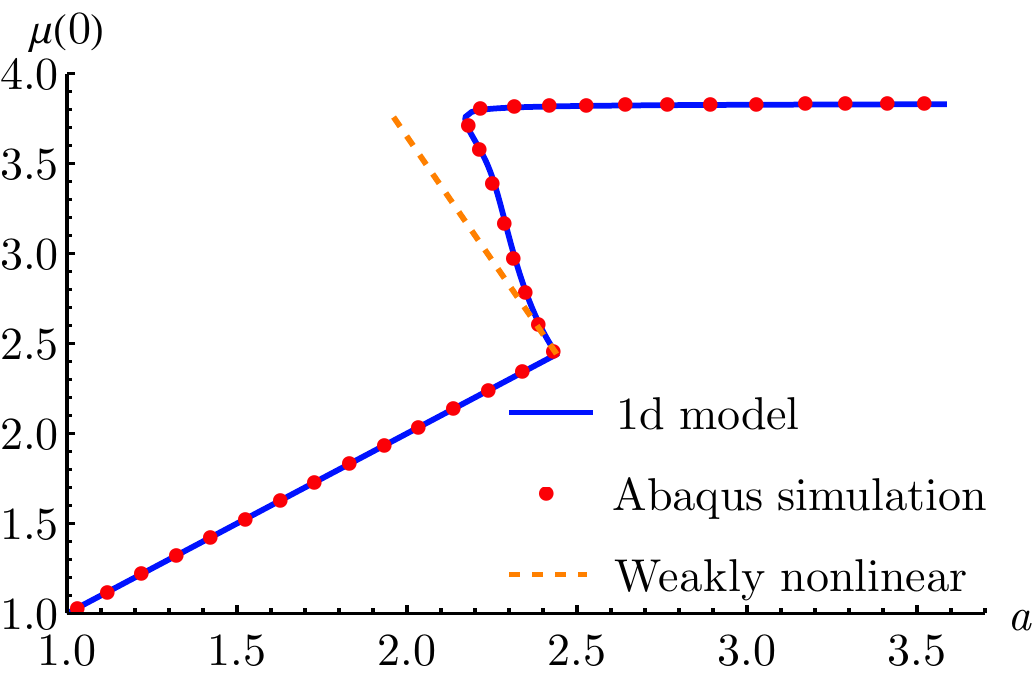}}
	\caption{Variations of two measures of necking amplitude with respect to the scaled radius predicted by  the 1d model, Abaqus simulations and weakly nonlinear analysis: (a) $\lambda(A)-\lambda(0)$ versus $a$ and (b) $\mu(0)$ versus $a$.}
	\label{fig:LC}
\end{figure}

\begin{figure}[h!]
	\centering
	\subfloat[]{\includegraphics[width=0.4\textwidth]{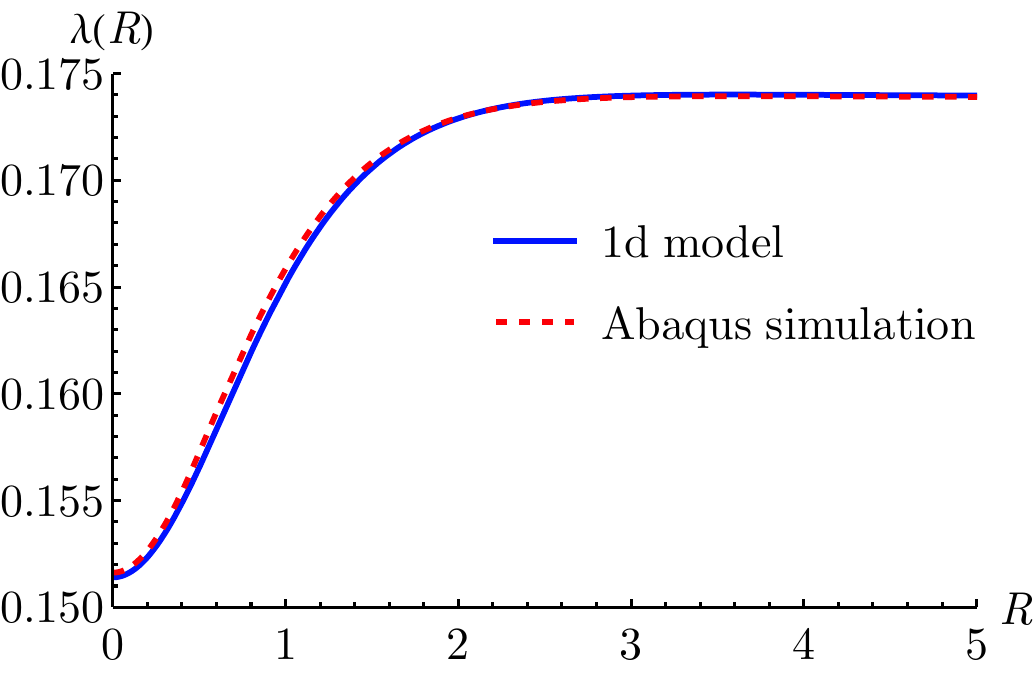}
	}\qquad\qquad
	\subfloat[]{\includegraphics[width=0.4\textwidth]{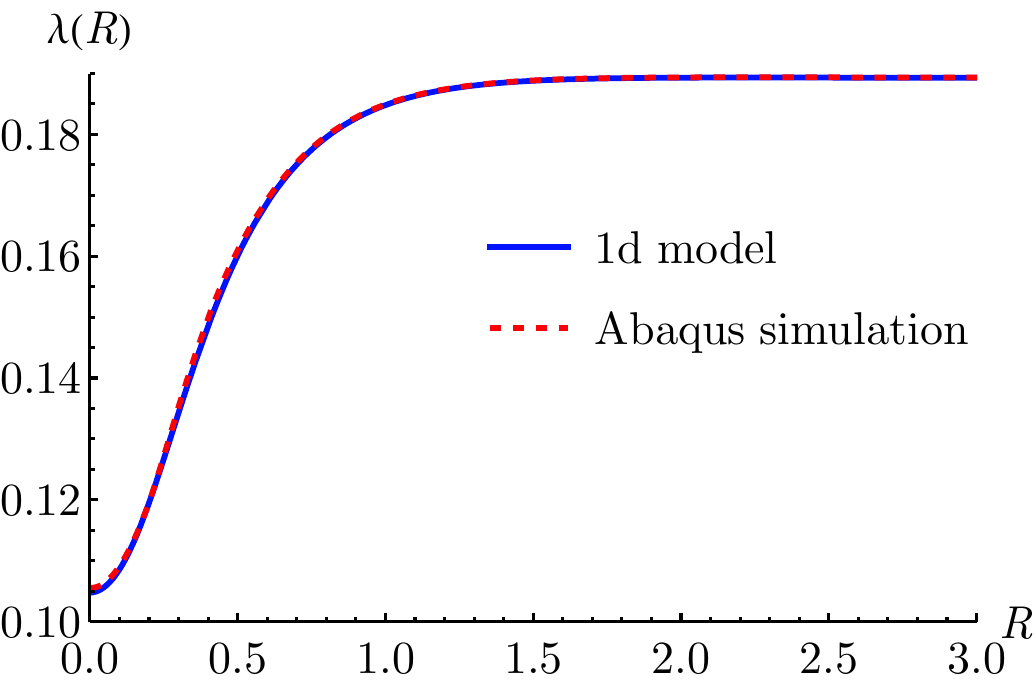}
	}\\
	\ \subfloat[]{\includegraphics[width=0.4\textwidth]{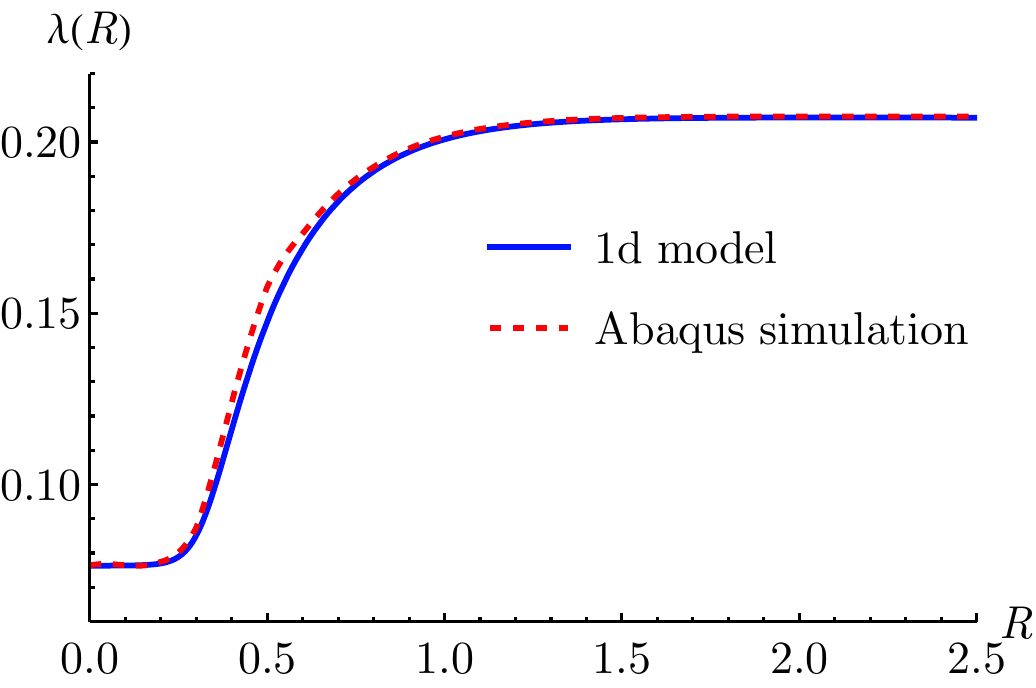}
	}\qquad\qquad
	\subfloat[]{\includegraphics[width=0.4\textwidth]{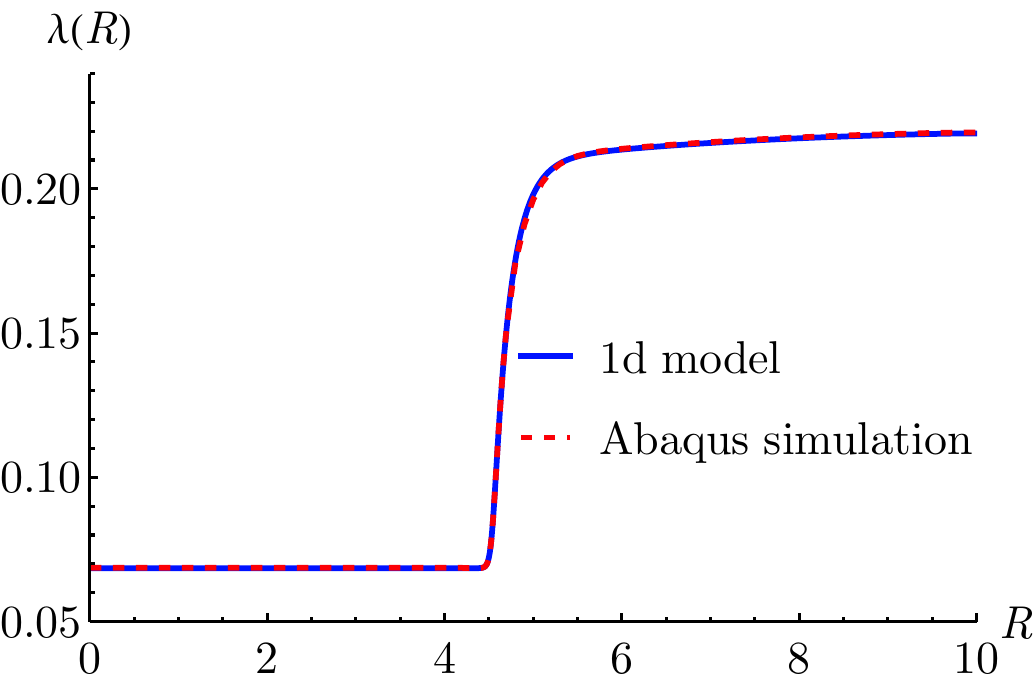}}
	\caption{Necking solutions based on the 1d model and Abaqus simulations  at the four states marked in Fig.~\ref{fig:LC}(a): (a) $a=2.4$, (b) $a=2.3$, (c) $a=2.2$, (d) $a=2.6$.}
	\label{fig:necking-solution}
\end{figure}

\begin{figure}[h!]
	\centering
	\includegraphics[width=0.58\linewidth]{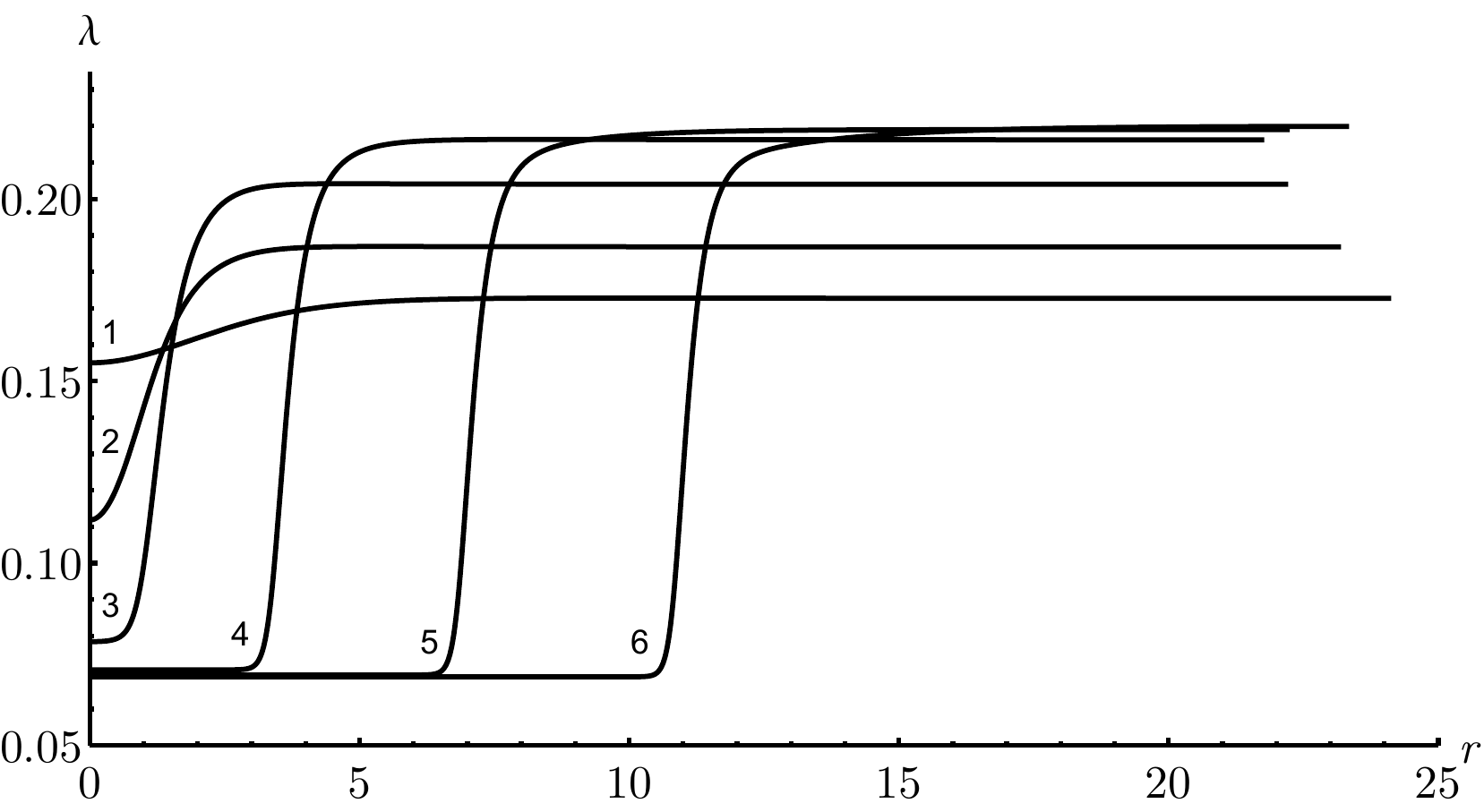}
	\caption{Evolution of the necking profile in the purely mechanical case predicted by the 1d model: the six curves correspond
		to $a=2.41$, $2.3$, $2.22$, $2.17$, $2.22$, and $2.37$, respectively, and $r=\mu(R)R$ refers to the current radial coordinate.}
	\label{fig:necking-evolution}
\end{figure}

In Fig.~\ref{fig:LC}, we show the variations of the two measures of necking amplitude, $\lambda(A)-\lambda(0)$ and $\mu(0)$, with respect to the scaled radius $a=\mu(A)=r(A)/A$. The necking solutions given by the 1d model and Abaqus simulations at the four states marked in Fig.~\ref{fig:LC}(a) are given in Fig.~\ref{fig:necking-solution}. It is seen that the 1d solutions agree well with Abaqus simulations in the entire post-bifurcation regime and the weakly nonlinear analysis captures the near-critical behavior correctly. Finally, Fig.~\ref{fig:necking-evolution} shows the evolution of the necking solution in the entire post-bifurcation regime. It is seen clearly from the figure that necking initially appears as a long wavelength mode, then grows locally to a maximum size, and finally propagates towards the edge.


\subsection{Maxwell characterization of necking propagation}

Like Maxwell's construction for phase separation, the propagation stage of necking can be characterized analytically using a simplified model. One observes from Fig.~\ref{fig:necking-evolution} that in the propagation stage, the regions where necking has or has not taken place each have a constant thickness stretch and are connected by a thin transition region. To simplify the analysis, let us assume that the thin transition region has zero thickness and refer to it as the interface. Also, for easy reference, let us refer to the thin inner region  as the ``$-$" phase and the thick outer region as the ``$+$" phase. We use the subscripts ``$-$" and ``$+$" to signify associations with the ``$-$" and  ``$+$" phases, respectively.

The ``$-$" phase (inner region) is homogeneous and its deformation is described by
\begin{align}
r=\lambda_-^{-1/2}R,\quad z=\lambda_-Z,\quad 0\leq R\leq B,
\end{align}
where $\lambda_-$ denotes the constant thickness stretch of  the ``$-$"phase and $B$ denotes the location of the interface.

For the  ``$+$" phase (outer region),  the incompressibility constraint implies that  the deformation is of the form
\begin{align}
r=\sqrt{\lambda_-^{-1}B^2+\lambda_+^{-1}(R^2-B^2)},\quad z=\lambda_+Z,\quad B \leq R\leq A,
\end{align}
where $\lambda_+$ is the constant thickness stretch of the ``$+$"phase. In particular, the azimuthal stretch of the ``$+$" phase is given by
\begin{align}\label{eq:mu+}
\mu_+=\frac{r}{R}=\frac{\sqrt{\lambda_-^{-1}B^2+\lambda_+^{-1}(R^2-B^2)}}{R}.
\end{align}

Since the thickness stretches are constants in the two phases, the total energy of the plate in the \lq\lq two-phase" state is obtained by neglecting the gradient term in \eqref{eq:1d} and is given by
\begin{align}\label{eq:Ep}
\mathcal{E}=\frac{1}{2}B^2 w(\lambda_-^{-1/2},\lambda_-^{-1/2})+\int_B^A w(\mu_+^{-1}\lambda_+^{-1},\mu_+)R\,dR-PA^2\mu_+|_{R=A}.
\end{align}
When  $P$ is given, $\mathcal{E}$ is a function of two thickness stretches $\lambda_-$ and $\lambda_+$, and the radius $B$ of  the interface. The stationary condition of $\mathcal{E}$ leads to
\begin{align}\label{eq:char}
\frac{\partial\mathcal{E}}{\partial\lambda_-}=0,\quad \frac{\partial\mathcal{E}}{\partial\lambda_+}=0,\quad \frac{\partial\mathcal{E}}{\partial B}=0.
\end{align}
The physical meanings of these equations are identified in \ref{app:propagation}. It is shown that the first equation corresponds to the continuity of the radial nominal stress across the interface, and the second equation represents the vanishing of the resultant on the upper/lower surface of the ``+" phase. The last equation can be viewed as an analogy of the well-known Maxwell equal-area rule. Once $\lambda_-$, $\lambda_+$ and $B$ are determined from \eqref{eq:char}, the necking solution in the propagation stage is well approximated by the following piecewise constant function:
\begin{align}\label{eq:lambdaR}
\lambda(R)=\begin{cases}
\lambda_-,&0\leq R< B,\\
\lambda_+,& B\leq R\leq A.
\end{cases}
\end{align}

\begin{figure}[h!]
	\centering
    \includegraphics[width=0.4\textwidth]{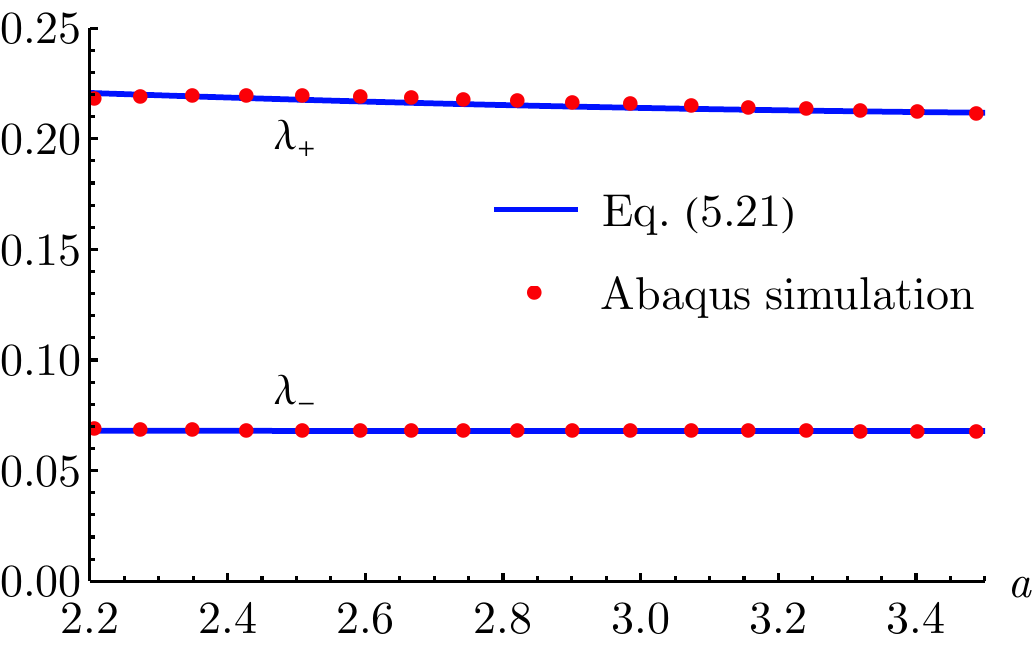}\qquad\qquad
    \includegraphics[width=0.4\textwidth]{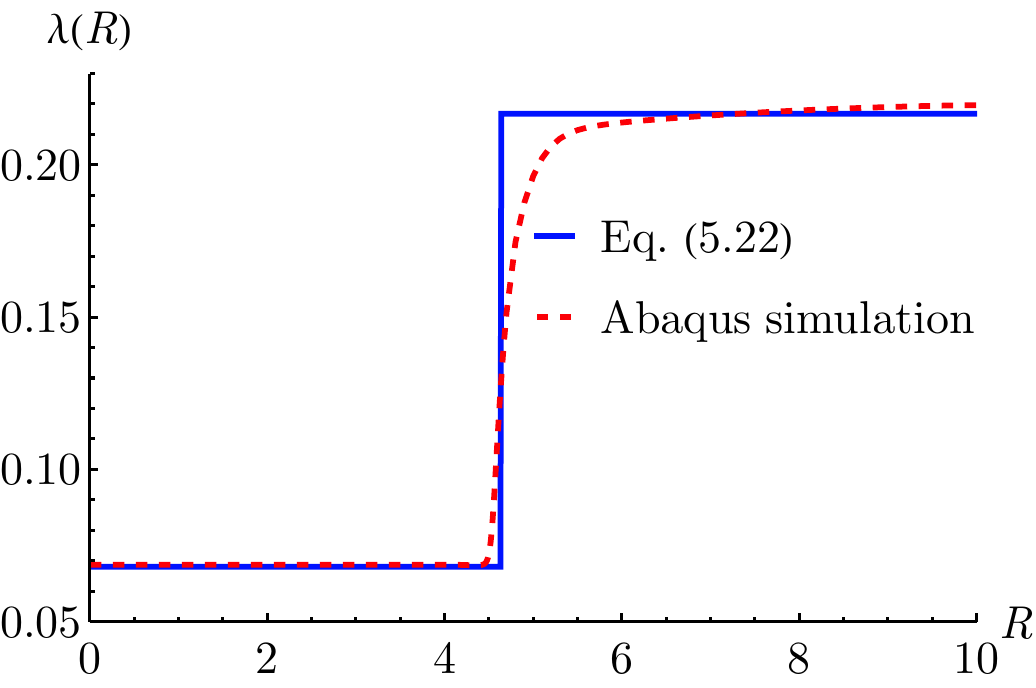}
	\caption{ (a) Thickness stretches of the thin and thick regions  given by Eq. \eqref{eq:char} and Abaqus simulations. (b) A typical necking profile, corresponding to $a=2.6$, given by Eq. \eqref{eq:lambdaR} and Abaqus simulations.}
	\label{fig:p}
\end{figure}

To quantify the accuracy of the simplified model, for the two-term Ogden material model \eqref{eq:mm}, we show in Fig \ref{fig:p}(a) the dependence of the thickness stretches $\lambda_-$ and $\lambda_+$ on the scaled current radius $a$ in the propagation state based on Eq. \eqref{eq:char} and Abaqus simulations. It is seen that there are good agreements in both $\lambda_-$ and $\lambda_+$. Fig.~\ref{fig:p}(b) displays a typical necking profile  predicted by Eq. \eqref{eq:lambdaR} and Abaqus simulations,  verifying that the former provides a good approximation for $\lambda(R)$.

\section{Application to axisymmetric necking of a dielectric plate}\label{sec:6}

Having validated the 1d model for the purely mechanical case, we now apply it to analyze the axisymmetric necking of a dielectric plate under equibiaxial stretching and an electric field. Two typical loading scenarios are considered: fixed edge \citep{koh2011mechanisms} or fixed tensile force \citep{huang2012giant}. We assume that the elastic behavior and geometry of the dielectric plate  are the same as that used in Section \ref{sec:5}, that is, with the strain energy \eqref{eq:mm} and $A=5H$.

\subsection{The case of fixed edge}
\begin{figure}[h!]
	\centering
	\subfloat[]{\quad\includegraphics[width=0.4\textwidth]{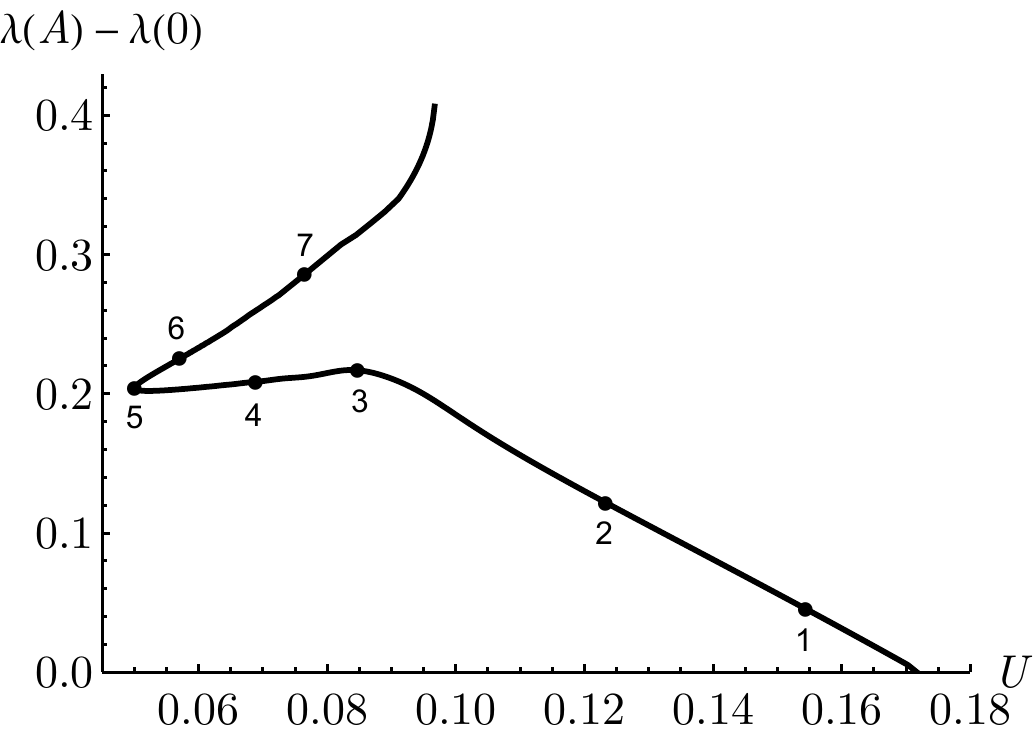}
	}\qquad\qquad
	\subfloat[]{\includegraphics[width=0.4\textwidth]{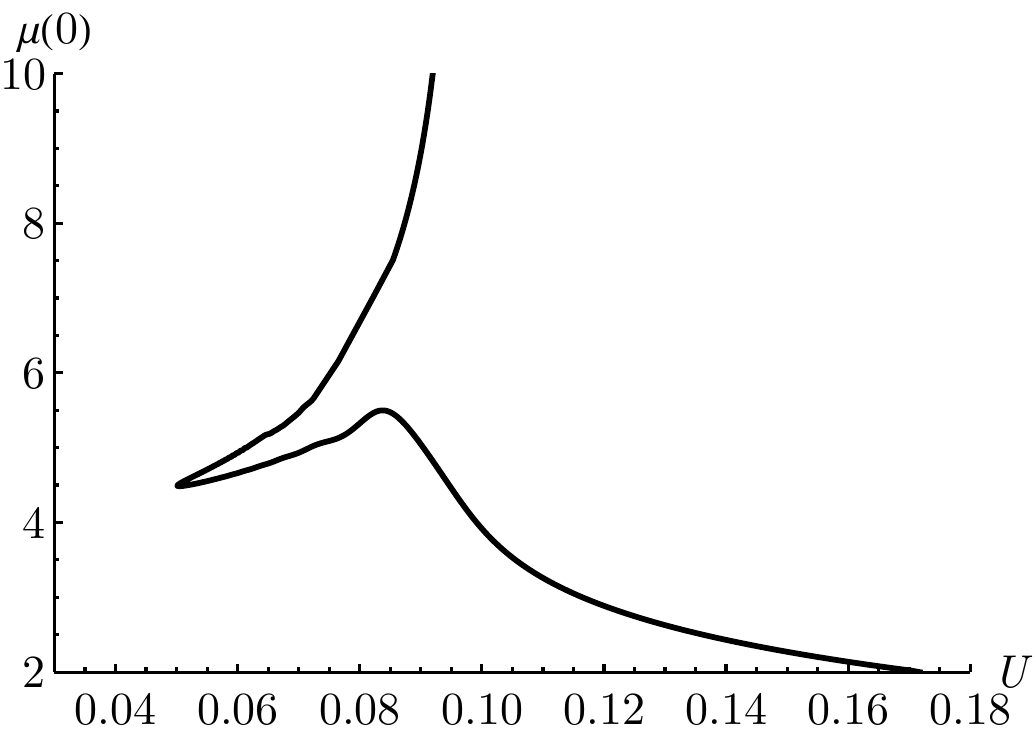}}
	\caption{Variations of two measures of necking amplitude with respect to the scaled voltage based on the 1d model: (a) $\lambda(A)-\lambda(0)$ versus $U$ and (b)  $\mu(0)$ versus $U$.  The edge of the plate is fixed with $a=2$.}
	\label{fig:LC-EM}
\end{figure}

\begin{figure}[h!]
	\centering
	\includegraphics[width=0.58\linewidth]{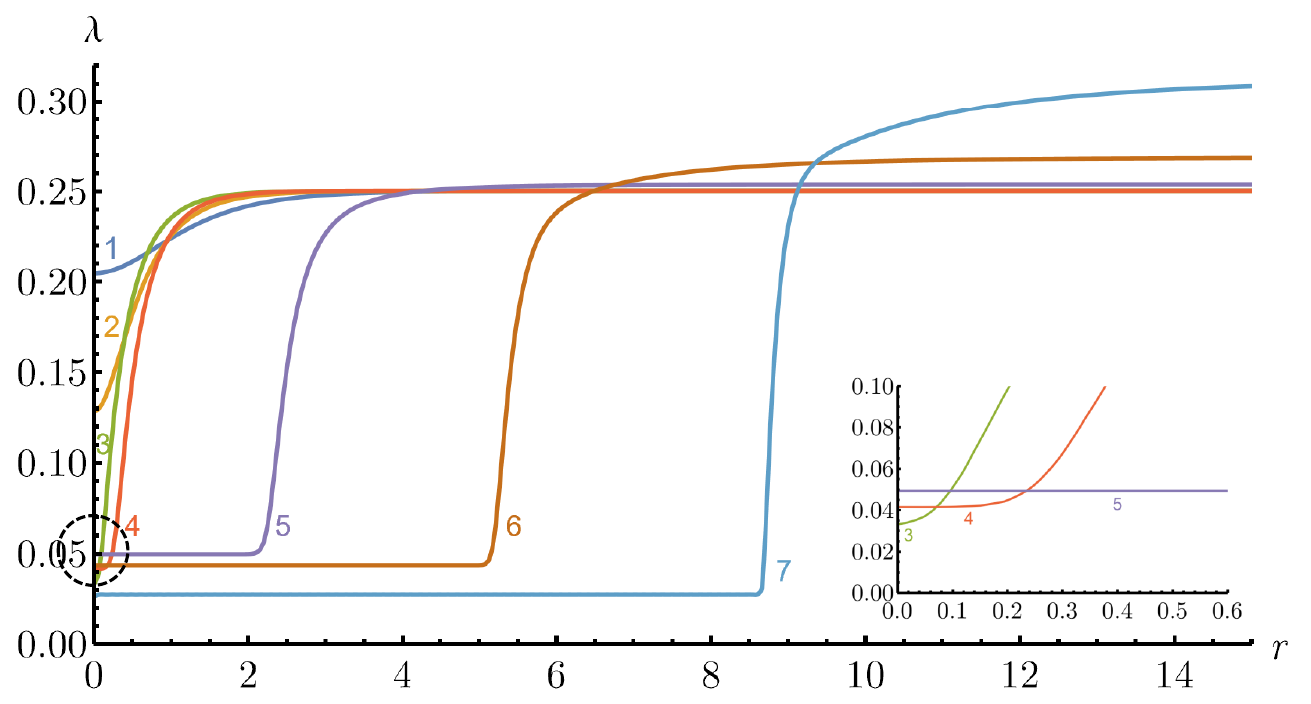}
	\caption{Evolution of necking profile of the dielectric plate predicted by the 1d model for fixed edge $a=2$. The small image is a magnified view of the circular region (marked by dashed lines), and $r=\mu(R)R$ refers to the current radial coordinate. The seven curves correspond to the states indicated in  Fig.~\ref{fig:LC-EM}(a), at which $U=0.154$, $0.123$, $0.085$, $0.069$, $0.05$, $0.057$ and $0.077$, respectively.}
	\label{fig:NE-fixed-edge}
\end{figure}

We first consider the case that the dielectric plate is first uniformly stretched, corresponding to a prescribed value of $\mu$ ($=r/R$), and then  the electric voltage is increased  from zero with the edge fixed
(so that $a=\mu(A)=\hat{a}$ is fixed when the deformation has become non-uniform). A linear bifurcation analysis shows that the bifurcation condition for axisymmetric necking in the
electroelastic case is given by \citep{fu2023axisymmetric}
\begin{align}\label{eq:bifur}
\mu^{-2}w_{11}(\mu,\mu)=U^2,
\end{align}
where we recall that $U$ ($=\sqrt{\epsilon}V/H$) is the scaled voltage. As an example, we take  $a=2$ (which corresponds to a thickness stretch $\lambda = 1/\sqrt{2}$).  The above condition then gives the critical voltage $U=U_\text{cr}=0.172$ at which necking initiates.
The subsequent necking evolution is determined with the use of the Rayleigh-Ritz method outlined in the previous section. Fig.~\ref{fig:LC-EM} shows the dependence of two measures of the necking amplitude, $\lambda(A)-\lambda(0)$ and $\mu(0)$, on the scaled voltage $U$; the necking solutions at the seven states indicated in Fig.~\ref{fig:LC-EM}(a) are shown in Fig \ref{fig:NE-fixed-edge}. It is seen from the figures that an inhomogeneous solution does indeed bifurcate from the homogeneous solution. It localizes rapidly and the amplitude $\lambda(A)-\lambda(0)$ grows until it has reached a maximum amplitude (states $1\sim3$). This phase of necking evolution is accompanied by gradual reductions in the electric voltage, which is the characteristic of subcritical bifurcations. With continued drops in voltage until it reaches a minimum (states $3\sim5$), the necking amplitude $\lambda(A)-\lambda(0)$ experiences a slight reduction.  After that, the necking propagates towards the edge with rapid rises in the magnitude (and thickness reduction at the origin); this propagation stage is maintained by increasing the voltage continuously (states $5\sim7$).
The reduction of necking amplitude corresponding to states $3\sim5$ is an exclusive feature of axisymmetric necking and is missing in the case of uniaxial stretching, where the amplitude always increases monotonically. This non-monotonic variation of amplitude is worth further
exploration.

\begin{figure}[h!]
	\centering
	\includegraphics[width=0.48\linewidth]{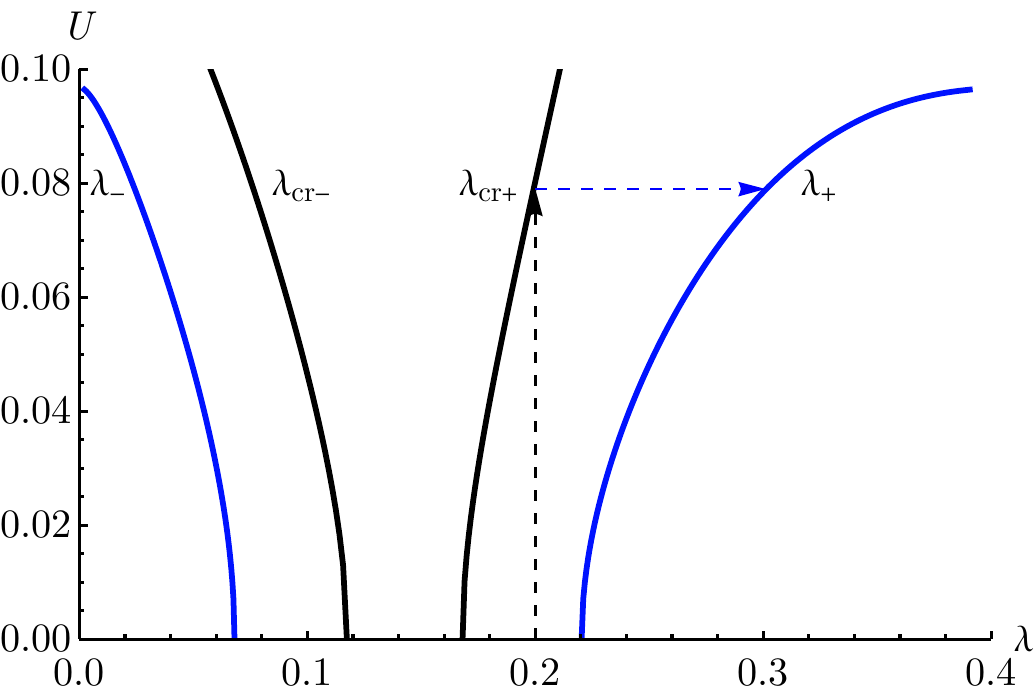}
	\caption{Variations of the four special values of $\lambda$ with respect to $U$.  The $\lambda_{\rm cr+}$ and $\lambda_{\rm cr-}$ are the two bifurcation values predicted by \rr{eq:bifur} with $\mu$ replaced by $1/\lambda^2$. The $\lambda_{+}$ and $\lambda_{-}$ are the values of $\lambda$ on the two sides of the interface in the propagation stage, determined by the counterpart of \rr{eq:char} for the electroelastic case.}
	\label{figadded}
\end{figure}

The growth stage (states $1\sim3$) and propagation stage (states $5\sim7$) of axisymmetric necking share similarities with their counterparts in the inflation of a rubber tube with fixed ends \citep{gwf2022} and surface-tension-induced necking of a solid cylinder with fixed ends \citep{fu2021necking}. Guided by the latter studies, we have presented in Fig.~\ref{figadded} the dependence on $U$ of the two values of $\lambda$ on the two sides of the interface in the propagation stage with $a=1/\sqrt{0.2} \approx 2.24$. For the current case, $\lambda_{+}$ is dependent on $a$, but this dependence is weak and can be neglected; see Fig.~\ref{fig:p}.  The vertical dashed line shows a typical loading path that starts from $\lambda=a^{-2}=0.2$ and $U=0$. When this vertical loading path intersects the curve labelled as $\lambda_{\rm cr+}$, axisymmetric necking would initiate. The necking solution can only be maintained by decreasing $U$, as shown in Fig.~\ref{fig:LC-EM}, and the propagation stage (states $5\sim7$) is characterized by the curves labelled as $\lambda_{+}$ and $\lambda_{-}$. If, after reaching the bifurcation point, $U$ were increased further above its bifurcation value, the membrane would snap through to a \lq\lq two-phase" state, as indicated by the blue (horizontal) dashed line in Fig.~\ref{figadded}, and further evolution of the propagation stage would follow the blue line. This is consistent with the fact shown in Fig.~\ref{fig:NE-fixed-edge} that in the propagation stage, $\lambda_{+}$ is increasing whereas $\lambda_{-}$ is decreasing. In this snap-through scenario, the membrane thickness at the origin would experience a drastic reduction determined by $\lambda_{-}$. This would almost certainly result in an electric breakdown if a breakdown had not already taken place earlier.

\begin{figure}[h!]
	\centering
	\includegraphics[width=0.5\linewidth]{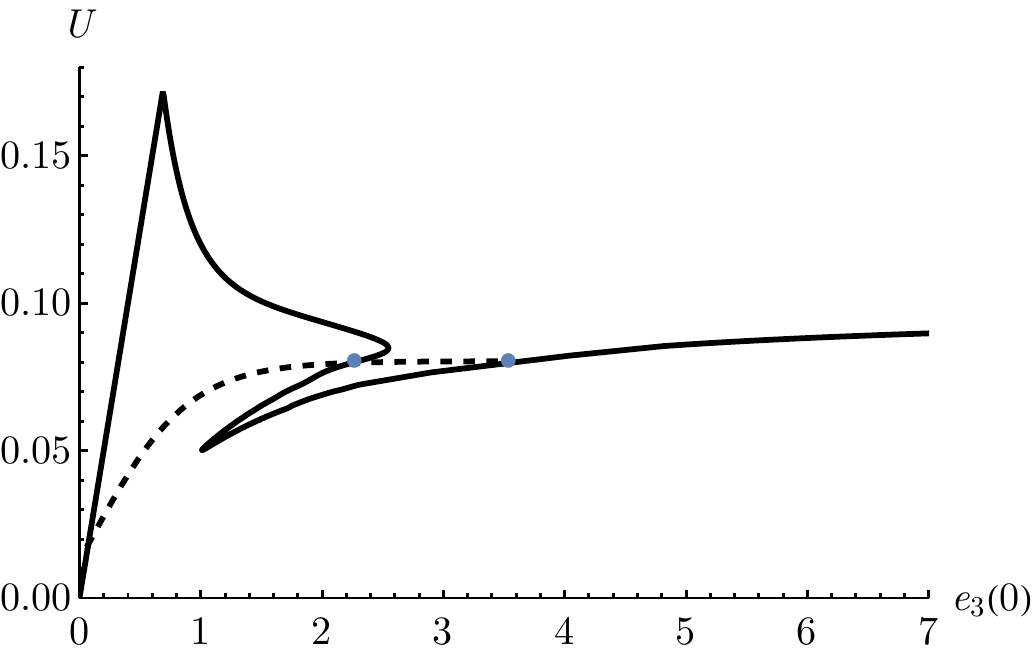}
	\caption{Relationship between the scaled voltage and true electric field $e_3(0)=U/\lambda(0)$ at $R=0$  with (dashed line) and without (solid line) imperfections for fixed $a=2$.}
	\label{fig:LC3-EM}
\end{figure}

As another scenario of electric breakdown,  we show in Fig.~\ref{fig:LC3-EM} the relationship between the  voltage and true electric field at $R=0$ for the entire loading process. The solid line represents the relationship for a perfect dielectric plate, while the dashed line shows a corresponding likely result for a dielectric plate with imperfections. This drop of loading curves  is similar to what happens in the buckling of spherical shells with imperfections \citep{hutchinson2016buckling}. For an in-depth exploration of how imperfections affect the loading curves, the reader is referred to \cite{fu2001nonlinear}.  If the evolution of the actual electric field were to follow the dashed line, then the snap-through to a \lq\lq two-phase" state leading to an electric breakdown would occur at a value of $U$ significantly smaller than its critical value (the maximum in Fig.~\ref{fig:LC3-EM}).

\subsection{The case of fixed tensile force}

\begin{figure}[h!]
	\centering
	\subfloat[]{\quad\includegraphics[width=0.41\textwidth]{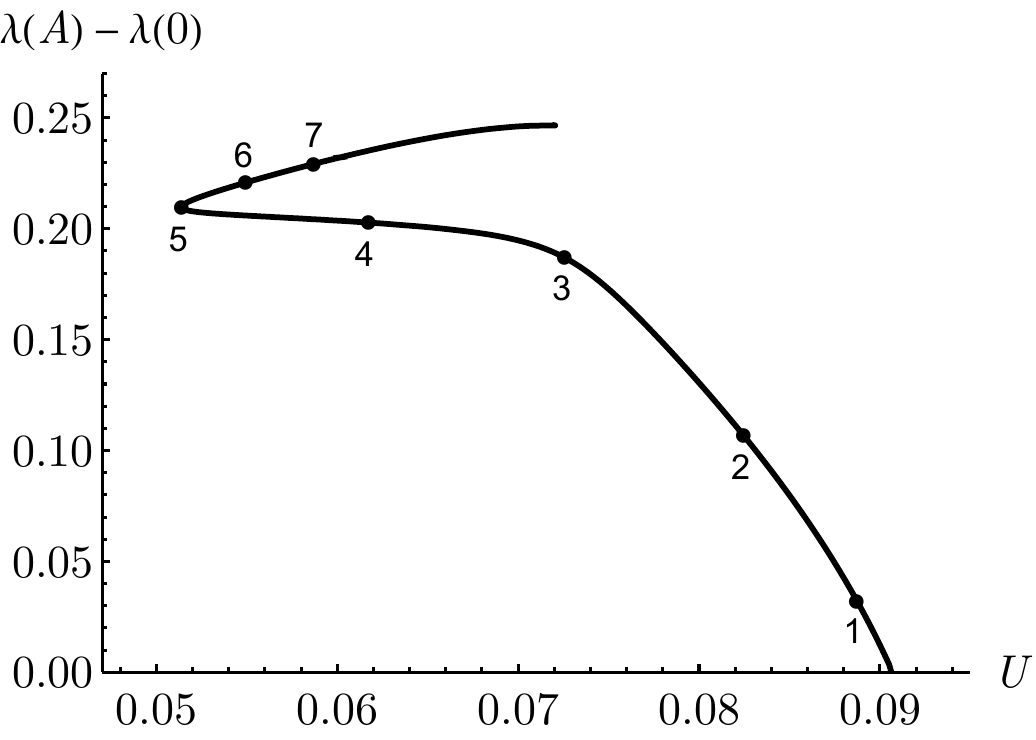}
	}\qquad\qquad
	\subfloat[]{\includegraphics[width=0.41\textwidth]{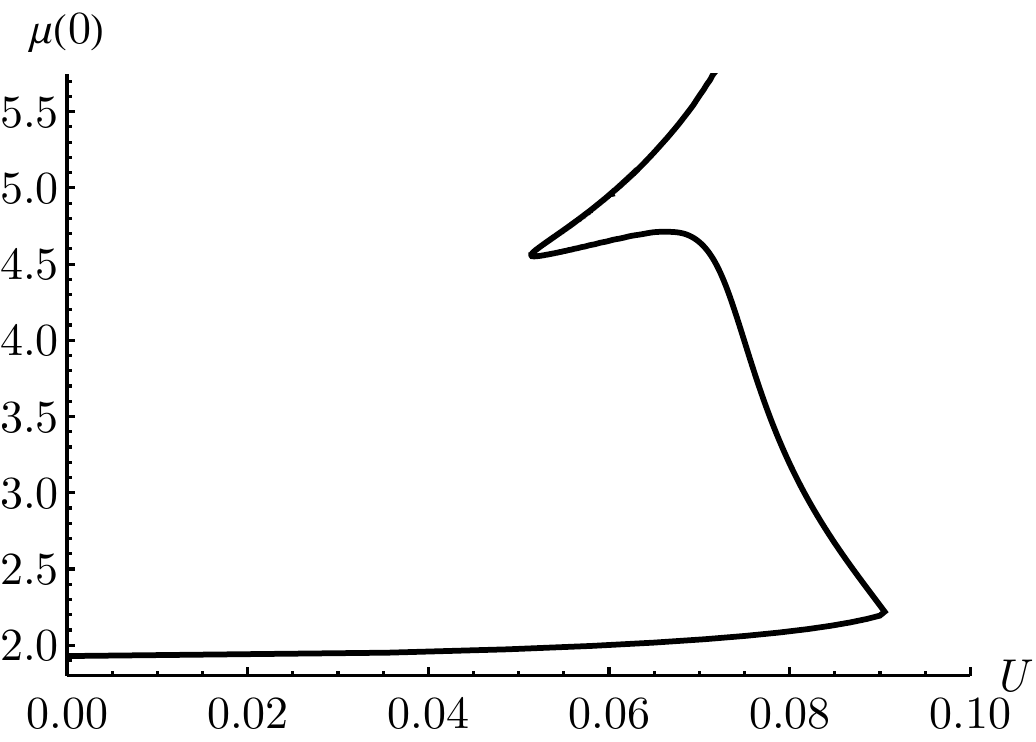}}
	\caption{Variations of two measures of necking amplitude  with respect to the scaled voltage based on the 1d model: (a) $\lambda(A)-\lambda(0)$ versus $a$ and (b) $\mu(0)$ versus $a$. The tensile force is fixed at $P=1.85$.}
	\label{fig:LC-P}
\end{figure}

\begin{figure}[h!]
	\centering
	\includegraphics[width=0.61\linewidth]{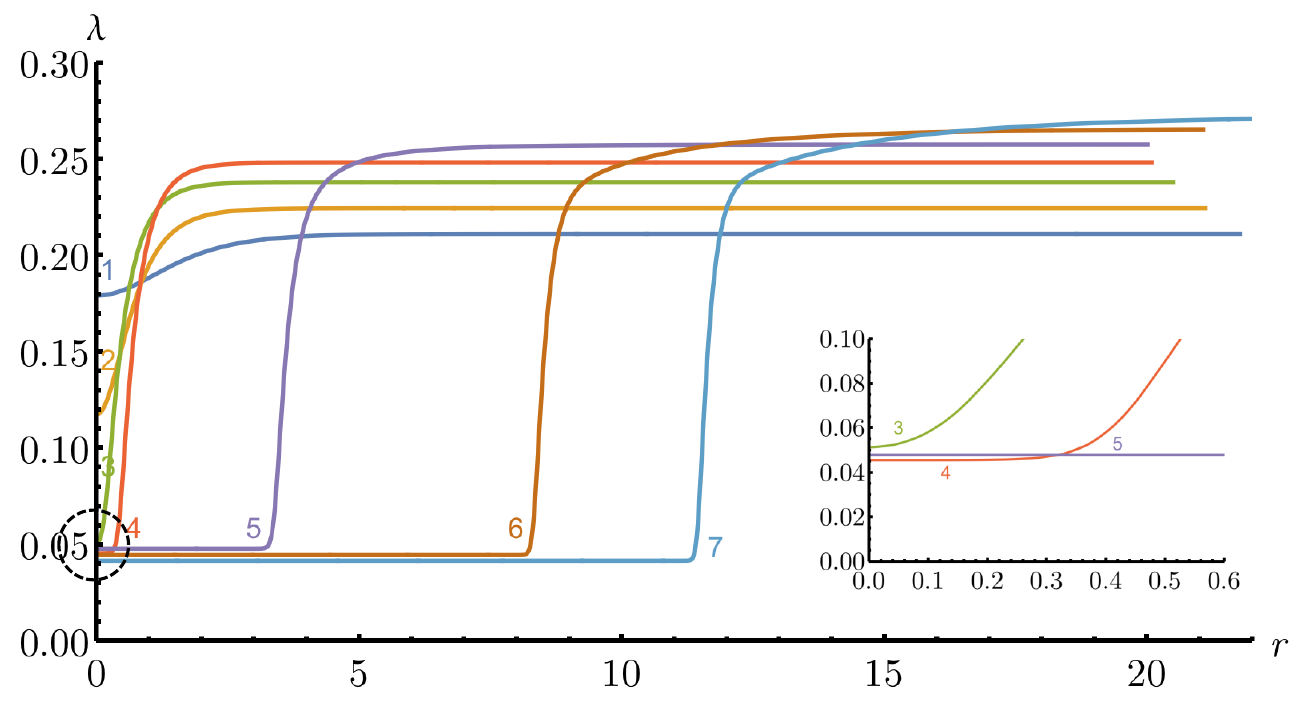}
	\caption{Evolution of the necking profile of the dielectric plate predicted by the 1d model for fixed tensile force $P=1.85$. The small image is a magnified view of the circular region (marked by dashed lines). The seven curves correspond to the states indicated in  Fig.~\ref{fig:LC-EM}(a), at which $U=0.089$, $0.082$, $0.073$, $0.062$, $0.051$, $0.055$ and $0.059$, respectively.}
	\label{fig:NE-fixed-P}
\end{figure}

In the second loading scenario, we consider the case when the tensile force $P$ is first increased to a specified value in the absence of the electric field, and then the electric field  is gradually increased/decreased with $P$ fixed.  In this case, it follows from \eqref{eq:P} and \eqref{eq:bifur} that the bifurcation values of $U$ and $\mu$ are determined by
\begin{align}\label{eq:bifP}
 \mu^{-2}w_{11}(\mu,\mu)=U^2,\quad w_1(\mu,\mu)- \mu^3 U^2=P.
\end{align}
We take $P=1.85$ as an example. Solving the two equations above then yields $\mu_\text{cr}=2.21$ and $U_\text{cr}=0.091$. As we trace the bifurcation solution away from the bifurcation point using the 1d model, the voltage drops while the necking amplitude $\lambda(A)-\lambda(0)$  grows until it has almost reached a maximum amplitude (states $1\sim 5$ in Fig.~\ref{fig:LC-P}), after which it propagates towards the edge with steady increases in both the necking amplitude and voltage.  The associated evolution of the necking profiles of the plate is shown in Fig.~\ref{fig:NE-fixed-P}.

\begin{figure}[h!]
	\centering
	\includegraphics[width=0.5\linewidth]{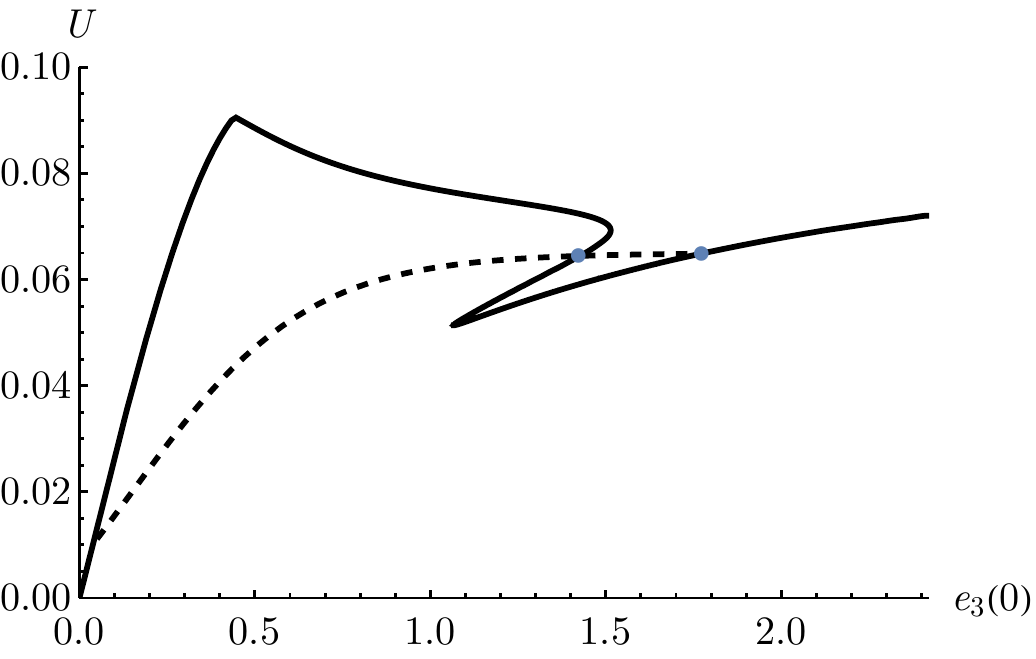}
	\caption{Relationship between the scaled voltage and true electric field at $R=0$ with (dashed line) and without (solid line) imperfections for fixed $P=1.85$.}
	\label{fig:LC3-P}
\end{figure}

In Fig.~\ref{fig:LC3-P}, we show the relationship between the  voltage and true electric field at $R=0$ for the entire loading process. One can see that the mechanism that leads to an electric breakdown in this case is similar to that in the fixed edge case. The presence of imperfections will reduce the critical voltage and a further increase of the voltage would trigger a snap-through instability that causes a significant rise in the true electric field. The sharp increase of the electric field may then lead to an electric breakdown of the dielectric plate.

\section{Conclusion}\label{sec:7}

We have developed a 1d model for analyzing axisymmetric necking in a dielectric membrane under the combined action of equibiaxial stretching and an electric field. The 1d model has been derived from the 3d nonlinear electroelasticity theory using the variational asymptotic method. The method constructed an asymptotic solution to the 3d variational problem  by taking advantage of the slowly-varying feature of the necking deformation: the leading-order solution is an adaptation of the homogeneous solution and the next-order solution is a small correction term included to maintain the self-consistency of the asymptotic expansion. As such, the resulting 1d model is asymptotically self-consistent, as evidenced by results in the linear and weakly nonlinear bifurcation analyses. To further appraise the range of validity of the 1d model, we have compared the predictions of the 1d model for axisymmetric necking with finite-element simulations, which reveals that the 1d model is highly accurate for characterizing axisymmetric necking in the entire post-bifurcation regime. We attribute the excellent performance of the 1d model to the self-consistency of the asymptotic solution.

The 1d model can be seamlessly integrated into numerical calculations thanks to its variational structure. This is particularly useful in the axisymmetric case, since it is unlikely to find analytic solutions due to the lack of translational invariance. In our illustrative examples, the 1d model is solved numerically using the Rayleigh--Ritz method, which is proved to be very efficient and robust. On the other hand, finite-element simulations using Abaqus are very likely to experience difficulties in the electro-mechanical case due to the complex, non-monotonic loading curves. It is because of the easy implementation and high efficiency of the 1d model that enables us to determine the evolution of axisymmetric necking completely in two commonly used loading scenarios. Thus, it is hoped that the methodology proposed in this paper will be useful in designing high-performance actuators that operate on the verge of instabilities.

Finally, we remark that although the 1d model presented is derived by assuming that the geometric and elastic parameters of the dielectric plate, as well as the applied voltage, are constant, it is possible to relax these assumptions and allow the geometric and electromechanical properties and the applied voltage to vary slowly in the radial direction. This includes the case of  dielectric plates with non-uniform thicknesses subjected to non-uniform voltages.

A Mathematica code that produces the results presented in Section \ref{sec:5} is available on GitHub (\url{ https://github.com/yfukeele}). It can be  adapted easily to generate the results in Section \ref{sec:6}.

\section*{Acknowledgements}

This work was supported by the National Natural Science Foundation of China (Grant Nos 12402068, 12072224), Guangdong Basic and Applied Basic Research Foundation (Grant No 2023A1515111141), and the Engineering and Physical Sciences Research Council, UK (Grant No EP/W007150/1).

\appendix

\section{Physical meanings of the equations characterizing necking propagation}\label{app:propagation}

In this section, we derive the explicit forms of the three equations in \eqref{eq:char}, and highlight their physical meanings.

First, in terms of the notations $\mu_-=\lambda_-^{-1/2}$ and $a=\mu_+|_{R=A}$, equation \eqref{eq:Ep} may be written in the form
\begin{align}
	\mathcal{E}=\frac{1}{2}B^2 w(\mu_-,\mu_-)+\int_B^A w(\mu_+^{-1}\lambda_+^{-1},\mu_+)R\,dR-PA^2a.
\end{align}
Applying differentiation under the integral sign, we obtain
\begin{align}\label{eq:Ea}
	\frac{\partial\mathcal{E}}{\partial\mu_-}=B^2 w_1(\mu_-,\mu_-)+\int_{B}^A (w_2 -w_1 \mu_+^{-2}\lambda_+^{-1})\frac{\partial \mu_+}{\partial\mu_-}R\,dR-PA^2 \frac{\partial a}{\partial\mu_-},
\end{align}
where $w_i$ $(i=1, 2)$ indicates the partial derivatives of $w$ with respect to $i$-th variable. From \eqref{eq:mu+} and $\lambda_-=\mu_-^{-2}$, we see that $\frac{\partial \mu_+}{\partial\mu_-}=\frac{\mu_-B^2}{\mu_+R^2}$ and consequently $\frac{\partial a}{\partial\mu_-}=\frac{\mu_-B^2}{a A^2}$. Hence \eqref{eq:Ea} may be reduced to
\begin{align}\label{eq:Eb}
B^{-2} \frac{\partial\mathcal{E}}{\partial\mu_-}=  w_1(\mu_-,\mu_-)-P\frac{\mu_-}{a}+\mu_-\int_{B}^A\frac{\mu_+ w_2-\mu_+^{-1}\lambda_+^{-1} w_1}{\mu_+^2 R}\,dR.
\end{align}

To simplify the integral in \eqref{eq:Eb} further, we observe that $\mu_+^{-1}\lambda_+^{-1}=\lambda_1$ and $\mu_+=\lambda_2$ are the principal stretches of the thick state in the radial and azimuthal directions. We also have
\begin{align}
\sigma_{11}-\sigma_{22}=\lambda_1 w_1-\lambda_1 w_2,
\end{align}
where $\sigma_{11}$ and $\sigma_{22}$ denote the radial and azimuthal  Cauchy stresses, respectively. Thus we can rewrite the integral in \eqref{eq:Eb} as
\begin{align}\label{eq:Ec}
\int_{B}^A\frac{\mu_+ w_2-\mu_+^{-1}\lambda_+^{-1} w_1}{\mu_+^2 R}\,dR=\int_{B}^A \frac{\sigma_{22}- \sigma_{11}}{\mu_+ r}\,dR,
\end{align}
where we have made the substitution $r=\mu_+R=\sqrt{\lambda_-^{-1}B^2+\lambda_+^{-1}(R^2-B^2)}$. With the use of the equilibrium equation
\begin{align}\label{eq:equi}
\frac{d\sigma_{11}}{dr}+\frac{\sigma_{11}-\sigma_{22}}{r}=0,
\end{align}
 and the relation $\lambda_{+} r dr=RdR$, we can reduce \eqref{eq:Ec} to
\begin{align}\label{eq:B7}
\int_{B}^A \frac{\sigma_{22}- \sigma_{11}}{\mu_+ r}\,dR=\int_B^A \frac{d\sigma_{11}}{dr}\frac{\lambda_+r}{\mu_+R}\,dr=\lambda_+\int_{A}^B d\sigma_{11}=\lambda_+\sigma_{11}|_{B}^A=Pa^{-1}-\lambda_+\sigma_{11}^+,
\end{align}
where use has also be made of the boundary condition $\sigma_{11}|_{R=A}=Pa^{-1}\lambda_+^{-1}$ and $\sigma_{11}^+=\sigma_{11}|_{R=B}$ denotes the radial Cauchy stress of the thick state at the interface. Finally, on substituting \eqref{eq:B7} into \eqref{eq:Eb}, we obtain
\begin{align}
B^{-2}\frac{\partial\mathcal{E}}{\partial\mu_-}= w_1(\mu_-,\mu_-)-\mu_-\lambda_+ \sigma_{11}^+.
\end{align}
It follows that $\frac{\partial \mathcal{E}}{\partial\lambda_-}=-\frac{1}{2}\lambda_-^{-3/2}\frac{\partial \mathcal{E}}{\partial \mu_-}=0$  is equivalent to
\begin{align}
 w_1(\mu_-,\mu_-)=\mu_-\lambda_+ \sigma_{11}^+,
\end{align}
which, in view of the fact that azimuthal stretch is continuous at the interface,  represents the continuity of the radial nominal stress across the interface.

A parallel calculation using differentiation under the integral sign yields
\begin{align}
\frac{\partial\mathcal{E}}{\partial\lambda_+}&=\int_B^A \lambda_+^{-1}\sigma_{33}R\,dR,\\
B^{-1}\frac{\partial\mathcal{E}}{\partial B}&=w(\mu_-,\mu_-)+ \sigma_{11}^+\Big(1-\frac{\lambda_+}{\lambda_-}\Big)-w(\mu_-^{-1}\lambda_+^{-1},\mu_-),
\end{align}
where $\sigma_{33}$ denotes the Cauchy stress in the thickness direction. The above equation clearly demonstrates that ${\partial\mathcal{E}}/{\partial\lambda_+}=0$ corresponds to the vanishing of the resultant on the upper/lower surface of thick state, and ${\partial\mathcal{E}}/{\partial B}=0$ is consistent with the jump condition derived in \cite{fu2004characterization}.

\bibliographystyle{model5-names}

\end{document}